\documentclass[10pt]{article}
\usepackage{mystyle,comment,xcolor}
\usepackage{quoting}
\usepackage{geometry}[0.9inch]
\usepackage{amsmath,amsthm,amssymb}

\definecolor{dgreen}{rgb}{0,.6 ,0}

\theoremstyle{definition}
\newtheorem{example}{Example}

\definecolor{newblue}{rgb}{.22,.37 ,.78}
\definecolor{neworange}{rgb}{.96,.61 ,.04}
\definecolor{newgreen}{rgb}{.18,.82 ,.43}

\begin{document}
\title{Don't let Ricci v. DeStefano Hold You Back: A Bias-Aware Legal Solution to the Hiring Paradox}  

\author{
\begin{tabular}{@{}c@{}}
Jad Salem \\
\normalsize \texttt{jsalem7@gatech.edu}
\end{tabular} 
\and
\begin{tabular}{@{}c@{}}
Deven R. Desai \\
\normalsize \texttt{deven.desai@scheller.gatech.edu}
\end{tabular} \and 
\begin{tabular}{@{}c@{}}
Swati Gupta \\
\normalsize \texttt{swatig@gatech.edu}
\end{tabular}
}

\date{\today}
\maketitle

\setcounter{page}{1}

\begin{abstract}
  Companies that try to address inequality in employment face a hiring paradox. Failing to address workforce imbalance can result in legal sanctions and scrutiny, but proactive measures to address these issues might result in the same legal conflict. Recent run-ins of Microsoft and Wells Fargo with the Labor Department's Office of Federal Contract Compliance Programs (OFCCP) are not isolated and are likely to persist. To add to the confusion, existing scholarship on Ricci v. DeStefano often deems solutions to this paradox impossible. Circumventive practices such as the 4/5ths rule further illustrate tensions between too little action and too much action. 
  
  In this work, we give a powerful way to solve this hiring paradox that tracks both legal and algorithmic challenges. We unpack the nuances of Ricci v. DeStefano and extend the legal literature arguing that certain algorithmic approaches to employment are allowed by introducing the legal practice of banding to evaluate candidates. We thus show that a bias-aware technique can be used to diagnose and mitigate ``built-in'' headwinds in the employment pipeline. We use the machinery of partially ordered sets to handle the presence of uncertainty in evaluations data. This approach allows us to move away from treating ``people as numbers'' to treating people as individuals---a property that is sought after by Title VII in the context of employment. 

\end{abstract}

\maketitle

\section{Introduction} 

How employers identify whom to interview and then hire has important effects across society. Employment significantly affects access to healthcare, continuing education and, therefore, quality of life. The benefits of employment are not, however, evenly distributed across race and gender categories in the United States. After George Floyd's death, companies acted to address racial injustice by making public statements, donations to support racial equality, and Juneteenth a company holiday \cite{duffy_2020}. Several companies went further. Microsoft announced a \$150 million investment to improve diversity including setting a goal of doubling the number of ``Black and African American people managers, senior individual contributors and senior leaders'' in the United States by 2025 \cite{duffy_2020_2}. Wells Fargo made a commitment to ``double Black Leadership'' by 2025 and ``will evaluate senior leaders based on their progress in improving diversity and inclusion in their areas of responsibility, in addition to other efforts'' \cite{duffy_2020_2}. Google has set a goal of having 30\% of its leadership from ``under represented groups'' by 2025 \cite{bass_2020}. Boeing seeks to increase representation of ``Black employees by 20\% while boosting other underrepresented groups over the next three years'' \cite{bass_2020}. Adidas announced plans to fill at least ``30\% of new positions with black or Latinx people'' \cite{duffy_2020_3}. Yet, both Microsoft and Wells Fargo received letters from the Labor Department's Office of Federal Contract Compliance Programs (OFCCP) due to concern that the plans may discriminate based on race \cite{duffy_2020}. At the same time, the OFCCP announced a settlement with Microsoft in September 2020 for \$3 million back pay and interest to address hiring disparities ``against Asian applicants'' for several positions from December 2015 to November 2018  \cite{U.S._Department_of_Labor}. The two OFCCP positions clash and appear to create a world where inaction opens the company to litigation, if not breaking the law, and corrective action creates the same risks. One might argue that the recent OFCCP inquiries were peculiar to the Trump administration's approach to this area of law and not something the current administration would pursue. Administrations, however, change and a new one might follow the Trump approach. Regardless of who is in the White House, legal activism to challenge steps taken to address diversity or challenge discriminatory results are not likely to go away. 

The reason this challenge is not likely to go away is that a company may be pursuing diversity goals and/or be addressing affirmative action plans; but the two are not the same, and the difference matters \cite{Estlund}. As the Equal Employment Opportunity Commission explains in ``Section 15 Race and Color Discrimination'' of its Compliance Manual, diversity can be understood as ``a business management concept under which employers voluntarily promote an inclusive workplace'' \cite{eeoc}. Companies have pursued diversity to attract talent and gain ``a competitive advantage'' \cite{eeoc}. In contrast, affirmative action refers to ``those actions appropriate to overcome the effects of past or present practices, policies, or other barriers to equal employment opportunity'' \cite{eeoc29cfr}. Such steps may occur because of a court order, negotiated settlement, or government regulation \cite{eeoc}. Employers may also use a voluntary affirmative action plan ``in appropriate circumstances, such as to eliminate a manifest imbalance in a traditionally segregated job category'' \cite{eeoc}. There is a conceptual and practical link between diversity goals and affirmative action. A company may pursue diversity ``for competitive reasons rather than in response to discrimination'' and ``such initiatives may also help to avoid discrimination'' \cite{eeoc}. As the legal status of diversity plans is unclear, methods to support both options are needed.

As another motivation, companies may want to see whether they are missing hiring and talent opportunities. Companies can be stuck in an equilibrium because they rely on, or exploit ``old certainties,'' rather than explore ``new possibilities'' \cite{March}. This exploration/exploitation trade-off began in organizational business literature but has become a significant part of how the machine learning community thinks about understanding information \cite{Domingos}. As a matter of best organizational and ML practices, companies need ways to explore new candidate pools. 

Regardless of the motivation behind a company plan, there is a steady drumbeat for algorithmic transparency, especially in employment and admissions contexts \cite{desai2017trust}. Thus an entity may have to or wish to reveal the process at some point. In either case the entity would want to show that their process is sound from both a mathematical and a legal view. These issues could push any company to avoid steps to address diversity because of litigation risks, both real and perceived. Although some scholars argue that the use of machine learning would constitute a valid business necessity claim so long as the target variable is job-related, thus rendering the question of equality of outcomes irrelevant, debates about which actions are and are not allowed to address diversity persist, especially when using an algorithmic approach \cite{bent2019algorithmic}.  
Simply put, when entities wish to be proactive regarding diversity, potential discrimination, or wish to explore whether they have missed opportunities in hiring talent \cite{kr18}, they will need a path that passes muster against a range of challenges.

This paper thus seeks to offer techniques and legal analysis to enable companies to pursue legal and ethical hiring goals and face this question: {\it How to improve equal opportunity and employment practices without crossing into arguably illegal discriminatory practices?} The ideas discussed here are general and key takeaways can be applied to several stages in the hiring pipeline. That said, this paper uses the screening stage of employment to exemplify methods and analysis and offer one way to attack the general problem.  

\section{Algorithms and the Hiring Process} 

\begin{figure}[t]
    \centering 
    \begin{tikzpicture}
    \draw[gray!20,fill = gray!20] (0,4) -- (0,-1.8) -- (-3.72,-1.8) -- (-3.72,4) -- cycle;
    \draw[gray!20,fill = gray!20] (7.44,4) -- (7.44,-1.8) -- (3.72,-1.8) -- (3.72,4) -- cycle;
    \draw[gray!20] (7.44,4) -- (7.44,-1.8) -- (-7.44,-1.8) -- (-7.44,4) -- cycle;
    \foreach \a in {0,...,9}{
        \foreach \b in {0,...,3}{
            \stickman{-7.1+\a*.33}{2.1+\b*.5}{0.1}{}
        }
    }
    \foreach \a in {0,...,9}{
        \foreach \b in {0,...,2}{
            \stickman{-7.1+3.72+\a*.33}{2.1+\b*.5}{0.1}{}
        }
    }
    \foreach \a in {0,...,9}{
        \foreach \b in {0,...,1}{
            \stickman{-7.1+3.72+3.72+\a*.33}{2.1+\b*.5}{0.1}{}
        }
    }
    \foreach \a in {0,...,9}{
        \foreach \b in {0,...,0}{
            \stickman{-7.1+3.72+3.72+3.72+\a*.33}{2.1+\b*.5}{0.1}{}
        }
    }
    \node at (0,.06){
    \begin{tabular}{p{3.3cm}p{3.3cm}p{3.3cm}p{3.3cm}}
        \hspace{.9cm} Sourcing & \hspace{1cm}Filtering & \hspace{.1cm}Hiring, Interviewing & \hspace{.9cm}Retention \\ \hline
        $\bullet$ Unequal job ad targeting & $\bullet$ Unnecessary barriers for some groups & $\bullet$ Implicit bias among employers & $\bullet$ Implicit bias among employers \\
         & $\bullet$ Reinforcing of societal inequalities via machine learning & $\bullet$ Unequal impact of background checks on the over-policed & $\bullet$ Workplace environment issues \\
          &   & $\bullet$ Referral bias &
    \end{tabular}
    };
    \end{tikzpicture}
    %}
    \caption{Some ethical concerns in various stages of the employment pipeline.}
    \label{fig:pipeline}
\end{figure}
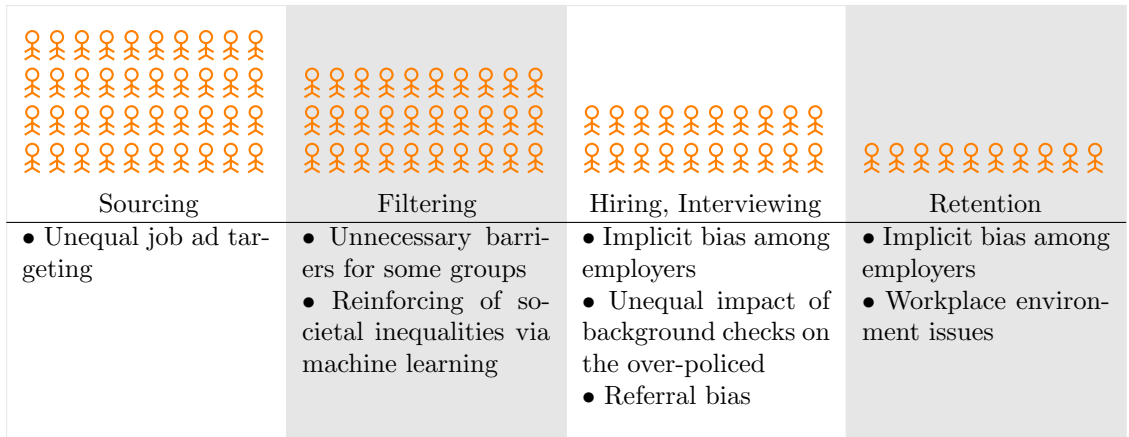

Employers want to hire a great workforce, but reaching and assessing the full viable range of potential employees poses problems. Many parts of the hiring process use algorithms as a way to manage and sort candidates. The practice can be traced back at least 40 years \cite{schwartz2019untold,shields2018over}. The problem is that there are a number of junctures in the hiring pipeline at which bias can affect decisions, as depicted in Figure \ref{fig:pipeline}. Job advertisements on various platforms can be targeted at specific audiences  \cite{angwin2017dozens,kim2020manipulating}. Application rates can differ across groups due to presumed employer bias \cite{mohr2014women}. Data-driven tools for evaluating r\'esum\'es can be biased due to inequalities in training data \cite{goodman2018why}, imbalance in data \cite{yucer2020exploring}, or differences in false positive/negative error rates in prediction algorithms leading to bias as a {\it downstream effect} \cite{dixon2018measuring}. Referral hiring can lead to favoritism \cite{schlachter2019employee}. Customer evaluations of freelancers can adversely impact certain groups \cite{hannak2017bias}. Final hiring decisions can be influenced by human biases of the hiring committee \cite{batastini2017bias}. After going through the hiring pipeline, candidates also see a significant difference in salaries offered \cite{yale}, and retention rates can differ dependent on the work environment \cite{glassdoor}. Indeed, societal biases are pervasive and can affect decisions made by experts \cite{hanks2009technology}. 

In addition, when automated systems are used at any stage, missed opportunity (false negatives) with respect to minority candidates is often shrugged off as an artifact of the prediction model, necessary for overall accuracy \cite{Kearns}. These models often train on historic data, which can depict imbalanced selection rates across different groups of candidates, and these trends can be learnt by automated methods \cite{barocas2016big,caliskan2017semantics}. History can dictate future actions. In short, existing pipeline practices can reiterate and increase disparity in opportunity and outcomes. Although the hiring pipeline can be improved in many places, we find the screening stage to be particularly ripe for improvement, and we therefore focus the article on this stage for the reasons outlined below. 

First, data-driven methods, by their nature, can pose a problem. Seemingly objective methods interact with real-world data, and so automated decisions can {\it reflect} and therefore, {\it reinforce} societal inequalities \cite{edelman2017racial,bogen2018help}. Even when there is no intent to discriminate, and the decision system uses the same data and applies the same rule to all, there may be a disproportionate effect on a protected class (i.e., groups protected by law from discrimination, such as those defined by sex, race, age, etc.) \cite{lum2016predict,barocas2014data,barocas-hardt-narayanan}. The problems in screening map to the more general ones present when using data-driven decision-making in hiring. So, screening is a good lens through which to investigate the concerns around using algorithms and data in the employment context in general. 

Second, algorithms are already used for screening applications. Such automated methods offer numerous advantages: speed, cost-effectiveness, potential objectivity, and uniformity in process. These properties may seem desirable at first glance from an ethical and fairness perspective; consistency in decisions is often a good thing, and a lack of human involvement would seem to minimize the role of implicit bias in hiring decisions \cite{Hanna_and_Linden}. Thus, automated methods have become commonplace in screening. Adjusting algorithmic techniques may be a more palatable idea and more feasible in an industry currently using automated processes than using algorithms in a heretofore un-automated process. New algorithmic interventions may, therefore, be more likely to be applied in practice.

Third, changes at early stages of the hiring pipeline are vital to address later bias. Changes at later stages are only meaningful if they act on a diverse pool of candidates. Without a diverse candidate pool at those stages, efforts to address bias become empty theater, because there will be few to no candidates from underrepresented groups for which the changes would help. As such, we focus specifically on automated screening processes: \emph{how should applicant-screening methods be developed?} These algorithmic tools should be designed with the goal of (a) selecting applicants of a desired quality, (b) satisfying some agreed upon fairness criteria, and (c) adhering to US anti-discrimination law.

\section{Biases in Data}  
\label{sec:bias-in-data}

We broadly refer to systematic inconsistencies in data which adversely affect certain groups as ``bias.'' The first step in reducing discrimination is to understand the source of this bias. Unfair decisions can stem from many places, and identifying the origins of the bias allows for precise interventions. In the hiring process (automated or otherwise), applications will typically be assigned a score, thus allowing comparisons of applicants based on a single number or with respect to a single ranking of candidates \cite{10.1145/3351095.3372849, jobsearch-rankings}. This evaluation metric can be hard-coded into an algorithm or developed dynamically, and in either case, can be unfair. A natural question is whether we can model this bias precisely and account for it within the algorithms to make them justifiably (provably) fairer.

Bias in evaluations can take different forms and be observed in different ways. For instance, a screening algorithm developed, {\it but not employed}, by Amazon penalized r\'esum\'es which included the word ``women's'' due to data of  past hiring trends in the company \cite{dastin_2018}. This algorithm penalized, for example, those who attended all-women colleges, and rewarded vocabulary typically used by men. In a similar vein, an empirical study showed that science faculty's assessment of r\'esum\'es varied dependent on the gender of the student \cite{yale}. These are fairly blatant examples of discrimination, as toggling a protected attribute results in different treatment. Note that this form of unfairness---while blatant---can be hard to observe in practice, as applicants are never truly identical but for a small number of attributes. 

Many cases of bias in evaluations, however, are more nuanced. Consider using SAT scores to screen candidates---a practice employers such as McKinsey, Bain, Goldman Sachs, and Amazon have been known to use even for candidates with advanced degrees \cite{dewan_2014, mckinsey, griswold_2014}. Studies show that even if students are equally able to perform well on a test, if the test is announced to exhibit differences across groups, students in a negatively stereotyped group perform lower than the students in a non-stereotyped group \cite{steele}. Another study from 2013 shows that SAT scores are correlated with family income, potentially pointing to issues of access \cite{SAT-Intersection}. Inside Higher Education looked at SAT scores in 2015 and found that despite fee waivers and increased efforts to provide support and tutoring to low-income families, 

\begin{quote}
``In each of the three parts of the SAT, the lowest average scores were those with less than \$20,000 in family income, and the highest averages were those with more than \$200,000 in income, and the gaps are significant. In reading, for example, the average for those with family income below \$20,000 is 433, while the average for those with income of above \$200,000 is 570.'' 
\end{quote}

\noindent Thus, compared to 2013, gaps in performance with respect to racial groups not only persisted but increased. This problem with SAT scores is further evident in a recent study by Faenza et al. \cite{faenza2020impact}, which showed a shift by approximately 200 points in SAT scores from schools with different {\it economic need indices}. Thus, an employer using SAT scores appears neutral but sets up a {\it pre-selected} pool. 

These issues regarding bias in data raise important \emph{design} questions for algorithmic intervention. When designing a decision-making algorithm, can we control for bias in historic data (thus avoiding Amazon's situation discussed above)? In other words, what steps can be taken to control for historic, economic, and/or social factors that are known to skew seemingly objective metrics such as the SAT?

\section{Approaches to addressing bias to date and their limits}
A variety of algorithmic techniques have been proposed for coping with biased data and improving fairness, from pre-processing techniques which involve modifying data before feeding it to an algorithm \cite{friedler2019comparative}; to in-processing techniques, which modify the algorithm itself \cite{kamishima2012fairness,zemel2013learning}; to post-processing techniques, which modify decisions made by an algorithm after the fact \cite{hardt2016equality,kamiran2010discrimination}. Current computer science literature highlights that merely scrubbing protected class information from an application may not help mitigate existing biases \cite{de2019bias}, and that algorithms have to use protected information to fix existing biases in data \cite{dwork}. Using protected information, however, may put the hiring process at odds with anti-discrimination law. Other prevalent approaches include iteratively removing data which is correlated with protected information \cite{zemel2013learning}; such approaches, however, may remove highly predictive information.

Algorithmic bias mitigation refers to the design of algorithms which perform well despite uncertainties about candidates' qualifications. This encompasses, for example, the design of procedures to select qualified candidates given biased data, or the design of algorithms which provably satisfy some notion of fairness. As discussed earlier, bias in evaluations can render bias-agnostic methods suboptimal \cite{kr18,emelianov2020on,closingthegap, dwork}; at the same time, imposing constraints such as demographic parity (i.e., proportional selection from different demographic groups) can hinder performance in some cases \cite{corbett2017algorithmic}, which points to potential trade-offs between bias mitigation and quality of selections. In our approach, we will take the view of algorithmic bias mitigation, given fine-tuned uncertainties in the evaluation of each individual.

\paragraph{Algorithmic Bias Mitigation.} Attempts to mitigate bias often begin with an understanding of the nature of the bias, or in other words, the inconsistencies in measurement of the ability of candidates. Mitigating the impact of such inconsistencies is an instance-specific endeavor; no cure-all exists. Nonetheless, there is theoretical work on mitigating bias under various mathematical assumptions. For instance, attempts have been made to address miscalibration of evaluations between multiple evaluators \cite{wang2019your}, and techniques have been developed for cases where some information is known about how biased each evaluator is in each evaluation \cite{wang2020debiasing}. In general, mathematical techniques can be developed as long as some assumptions on bias are made.\footnote{This points to multiple issues in bias-mitigation. First, the assumptions on bias are difficult to justify empirically, as ``ground truth'' is seldom available (for example, the true ability of a candidate is never truly known, especially for candidates who are not hired). Second, it is difficult to assess bias-mitigation techniques for a similar reason: if one does not know the ground truth, then it is hard to quantify how good any decision is.} Certain ``coarse'' sources of bias seem to be prevalent across demographic groups, and algorithms can be designed with these in mind. One might say these are the first approximations to incorporate the knowledge of large trends visible broadly across demographic groups, such as are seen in SAT scores discussed earlier \cite{faenza2020impact}. Addressing these coarser sources of bias from a theoretical point of view can provide insight in dealing with other forms of bias. 

A recent mathematical model that captures the dependence of errors in testing over groups is the \emph{group model} of bias. The model is based on the empirical work of Wenner{\aa}s and Wold \cite{wold1997nepotism}, and was introduced by Kleinberg and Raghavan in the context of offline selection (e.g., applicant-screening) \cite{kr18}, further studied by Salem and Gupta \cite{closingthegap}, Faenza et al. \cite{faenza2020impact}, and Blum and Stangl \cite{blum2020recovering} in the context of selection problems. This model assumes that bias is fairly consistent within each demographic group, and thus evaluations offer more accurate rankings within each group, but not across the groups. For example, once one accounts for difficulties in comparing one demographic group to another, there may be no way to confidently compare an 90\% attained by a white male scholar Adam to a 85\% attained by a Latina scholar Tia. But one can compare Adam against another white male scholar John with 83\%, and note that Adam is better. 

This model is at the same time appealing and dissatisfying in its simplicity. It is appealing in the sense that the model sheds light on best practices when the data is biased consistently for certain groups. That consistency indicates that information about group membership alone allows selection algorithms to reduce bias in selections. It is dissatisfying, however, in its coarseness, as it ignores intra-group differences in testing/evaluation errors and ignores any potential comparisons between groups. Adding to the example above, let us say that Tia also belongs to a low-income family, and we want to compare Tia to another Latina %Latinx female 
scholar May (not from a low-income family). This model does not account for such confounding variables of socio-economic status. Follow-up work by Celis et al. \cite{celis2020interventions} proposed a multiplicative model of bias in the context of rankings, wherein candidates in the intersection of different groups face a consistently higher bias. This approach, however, again equalizes the amount of bias within each smallest ``unique'' group (e.g., male, white, and age above 45 or lesbian, Asian, aged 39). It may not be okay to equalize the experience of every male, white person above the age of 45 or of every lesbian, Asian, under age 50. The underlying problem with this is the assumption of group membership, which may not even be accurate in practice. Indeed, whether a Chinese Asian, an Indian Asian, and a Filipino Asian faces the same amount of bias, and so should be treated the same, seems unlikely.

\paragraph{Current Industrial Practices.} How then do companies actually hire candidates, while reconciling with anti-discrimination laws and biases in the hiring pipeline? In a recent survey, the only specific public claim made by vendors of pre-employment assessments was adherence to the 4/5ths rule---outlined in the 1978 Uniform Guidelines on Employee Selection Procedures---which requires that group-specific selection rates of any pre-screening are all within a factor of $4/5$ of each other \cite{raghavan2020mitigating}. Yet this approach is coarse as it is agnostic to quality of candidates. Applying a 4/5ths rule in selection up front (e.g., as the current practice in the industry suggests \cite{raghavan2020mitigating}) does not change the perceived potential of candidates, nor account for uncertainties and biases in the data systematically. It can therefore simply set up the underrepresented group's candidates for failure, and lead to resentment and enlivening of negative stereotypes \cite{fischer2007effects,heilman1997affirmative}.

The trade-offs in algorithmic approaches track legal issues. If an employer uses an algorithmic tool to evaluate and screen candidates, the employer may face legal challenges depending on the outputs of the tool. A likely challenge is that the tool created illegal disparate impact. Disparate impact addresses when ``facially neutral policies or practices have a disproportionate adverse effect or impact on a protected class” \cite{ftc-big-data-report,accord-42}. The disparate impact doctrine is thus supposed to address situations where intent is not at hand or cannot be ascertained \cite{sullivan2005disparate}. In short, outcomes based on unaware algorithms may fit quite well with disparate impact challenges, because unaware algorithms are facially neutral, may lack intent to discriminate, and nonetheless yield statistically discriminatory results.\footnote{Despite the fact that the 4/5 rule is mentioned as evidence of disparate impact in the 1978 Uniform Guidelines on Employee Selection Procedures, there is no precise quantification of disparate impact. The 4/5 rule is often used as a trigger for litigation, but other statistical tests have been used in courts as well \cite{miao2013properties,sobol1988measures}.} 

The possibility of a disparate impact claim leads to an obvious approach. An employer may design a more aware algorithm that takes protected class status into account. However, this approach may run into a disparate treatment challenge. Disparate treatment is the legal doctrine that prohibits intentional use of race or other protected classes in making an employment decision. Thus, we return to the paradox described above, because it seems that an employer is trapped between using facially neutral systems that reflect systemic and historically conditioned, biased results or facing lawsuits for using aware systems to mitigate such effects. This paradox is exacerbated by current legal scholarship debating what algorithmic interventions to address bias, if any, are allowed and the implications of the lawsuit {\it Ricci v. DeStefano}, in which an action by the City of New Haven that tried to account for disparate impact of an administered promotion test led to litigation that was decided against the city. In Section \ref{sec:new-directions}, we outline the \emph{poset approach}, which we argue provides a way to solve the hiring paradox. Section \ref{sec:law-4-8-21} turns to an in-depth discussion on the takeaways from {\it Ricci v. DeStefano} and explains how the poset approach fits within legal rules so that one can use a bias aware approach to hiring and yet maintain individualized assessments of candidates.   

\section{A new approach: coping with uncertainty using partial orders}
\label{sec:new-directions}

\begin{figure}
    \centering
    \includegraphics[width=\textwidth]{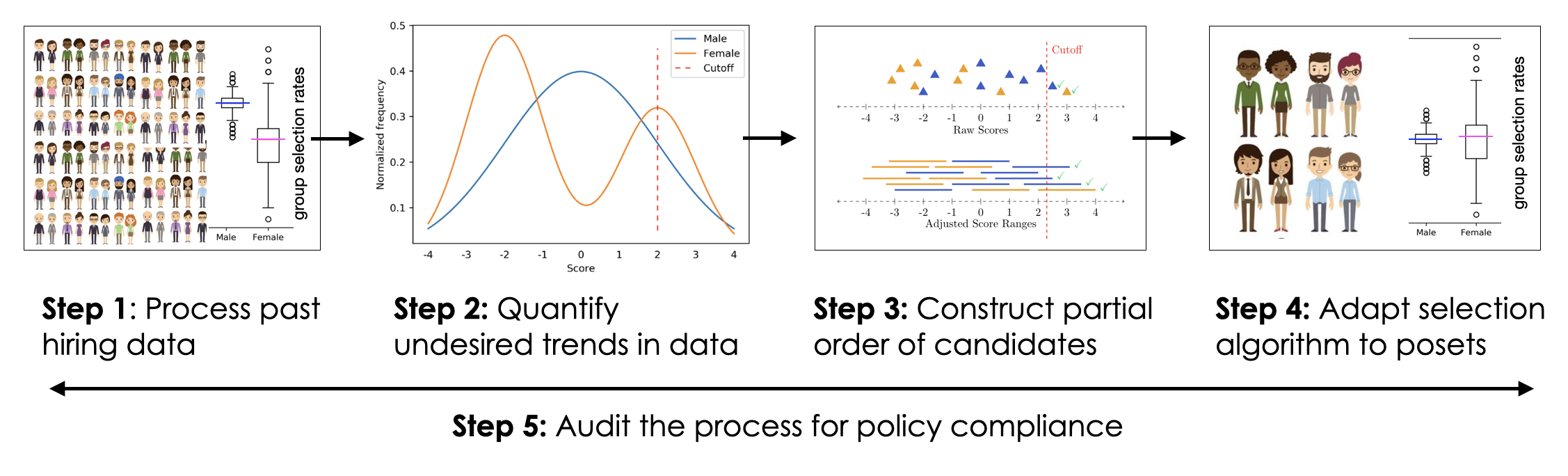}
    \caption{Coping with Uncertainty using Partial Orders: A Path to Address Disparate Impact in Hiring Practices}
    \label{fig:overview}
\end{figure}

As discussed in Section \ref{sec:bias-in-data}, coping with uncertainties in data is a fundamental problem in applicant screening systems, as well as in data-driven decision-making more generally. In this section, we will discuss one method, called the \emph{poset approach}, for applicant-screening in the face of uncertainty which has emerged recently in the computer science literature \cite{closingthegap}. In Section \ref{sec:law-4-8-21}, we will use this approach as a vehicle for discussing the legality of algorithmic bias mitigation in hiring. 

Consider the following scenario: there are three candidates $A$, $B$, and $C$, with ability scores of 82, 68, and 67, respectively, and you wish to grant interviews to two of them. The ability scores are known to be a strong predictor of job performance, but are only known to be accurate up to 3 points. In this case, there is a significant chance that $C$ is a better candidate than $B$, but the utilitarian approach of selecting the highest-scoring candidates would routinely select $A$ and $B$. The core idea behind the poset approach is that the latter approach is unfair to $C$, or more generally, that ignoring uncertainty can result in unfair decisions. In other words: 

\begin{quoting}[vskip=8pt]
    \it
    \noindent Some applicants, due to individual experiences or lack of historic data, cannot be reliably ranked. The solution need not involve producing a (possibly inaccurate) ranking. Instead, allowing for partial rankings can itself open the door to fairer decisions.
\end{quoting}

The poset approach, which we explain in more detail below, makes use of a mathematical structure called a \emph{partially ordered set}, or \emph{poset}, which can be used to encode uncertainty in ordinal information. Consider, for example, a set $S_1=\{4,2,5\}$ of true hirability of three candidates (which is often not observable in practice). This set is called \emph{totally ordered} since any pair of the scores can be ordered (i.e., ranked) with respect to the relation $\leq$. In other words, we can rank the scores: $2 \leq 4 \leq 5$, thus inducing an order amongst the candidates. 

However, in practice, one cannot observe directly how good a candidate might be at their job. This is where the poset approach can help. Intuitively speaking, one can think of a partial order as a set of comparisons, which may not cover all pairs of candidates (i.e., a total order with some comparisons missing). For example, consider a candidate $A$ who has experience in industry, a candidate $B$ who has experience in industry {\it and} who has an MBA, and a candidate $C$ who has an MBA. Considering these traits as binary (yes/no) attributes, one can represent their qualifications as the set $S_2 = \big\{\{\mbox{industry}\},\{\mbox{MBA}\},\{\mbox{industry},\mbox{MBA}\}\big\}$. From the given information, one might rank $B$ above both $A$ and $C$, since $B$ is qualified with respect to both measures, and the other candidates are only qualified with respect to one. However, $A$ and $C$ might be considered \emph{incomparable}, since neither candidate's qualifications subsume the other's. In this case, $S_2$ is a partially ordered set, but not a totally ordered set. To be precise, a relation $\preceq$ is a partial order on a set $S$ if three conditions hold for all $a,b,c \in S$: (1) $a \preceq a$; (2) if $a \preceq b$ and $b \preceq c$, then $a \preceq c$; and (3) if $a \preceq b$ and $b \preceq a$, then $a = b$. One can check that all these properties are satisfies for the set $S_2$. A poset is often visually depicted using its \emph{Hasse diagram}, which is directed graph in which edges represent orderings. For example, the Hasse diagrams for $S_1$ and $S_2$ are as follows:
\begin{center}
\begin{tikzpicture}[transform shape]
\tikzset{>=latex}
\node at (-1,1) {$S_1$:};
\node (2) at (0,.2) {2};
\node (4) at (0,1) {4};
\node (5) at (0,1.8) {5};
\node at (3,1) {$S_2$:};
\node (1s) at (4,.2) {$\{\mbox{industry}\}$};
\node (2s) at (5.5,.2) {$\{\mbox{MBA}\}$};
\node (12s) at (4.75,1.8) {$\{\mbox{industry, MBA}\}$};
\draw [->] (2) -- (4);
\draw [->] (4) -- (5);
\draw [->] (1s) -- (12s);
\draw [->] (2s) -- (12s);
\end{tikzpicture}
\end{center}
Note that Hasse diagrams omit redundant edges: even though $2 \leq 5$, the edge $2 \to 5$ is not included, since it is implied by the edges $2 \to 4$ and $4 \to 5$.

The \emph{poset approach} is the process of (1) forming a partial ranking (i.e., a partial order) of the candidate pool based on uncertainties, inaccuracies, or biases in data, and (2) making selections based on this poset. By making selection decisions in this way, one can concretely take uncertainty into account and, say, avoid routinely harming candidate $C$ in the example above. This can lead to bias mitigation in cases where the evaluation metric is biased against a certain group; e.g., if a group is underrepresented in training data and experiences large errors in the resulting ML model, the poset approach can confer benefit of the doubt to those underrepresented candidates.

We next illustrate how posets can model uncertainty using two examples.

\begin{example}
\label{ex:gender}
The poset approach can illustrate how one can account for uncertainties while also avoiding prohibiting discrimination based on gender.\footnote{As recently as June 15, 2020, the Supreme Court of the United States ruled that the Title VII of the Civil Rights Act prohibits discrimination on the basis of sexual orientation and gender identity \cite{bostock,npr-article}.} Using the poset approach, one may incorporate demographic context of the candidates and quantify uncertainy in their evaluations (either by directly observing the context, or by unsupervised methods such as clustering). Suppose that in a training dataset, nonbinary candidates are underrepresented, and as a consequence have high variance in errors in the prediction model. One may find that a nonbinary candidate Max has a wide score range of 80-90\% (e.g., due lack of training data on nonbinary candidates), another male candidate Adam has a score between 85-87\%, and a third female candidate Trisha has a score between 92-95\% (see Fig. \ref{fig:gender-example}). Now, using only the score ranges to compare candidates, Trisha compares favorably to Max, but it is unclear if Max is more qualified than Adam as their ranges overlap. In this case, we can think of Max and Adam as mutually incomparable. The poset approach therefore allows for \uline{individualized treatment} of inconsistencies in data processed. 

\begin{figure}
    \centering
    \begin{tikzpicture}
    % \draw[dashed,<->] (3,-.7) -- (11,-.7);
    \foreach[evaluate={\b=int(80+5*\a)},evaluate={\c=1.5*\a+3}] \a in {0,...,3}{
    \draw (\c,-.8) -- (\c,-.6);
    \node at (\c,-1.08){\b};
    }
    \draw[dashed,<->] (1.5,-.7) -- (9,-.7);
    
    \draw[ultra thick] (6.6,-.3) -- (7.5,-.3) node [midway, above=3pt, fill=white] {Trisha};
    \draw[ultra thick] (4.5,0) -- (5.1,0) node [midway, above=3pt, fill=white] {Adam};
    \draw[ultra thick] (3,-.3) -- node [fill=white]{Max} (6,-.3);
    
    \draw[ultra thick] (6.6,-.4) -- (6.6,-.2);
    \draw[ultra thick] (7.5,-.4) -- (7.5,-.2);
    \draw[ultra thick] (4.5,-.1) -- (4.5,.1);
    \draw[ultra thick] (5.1,-.1) -- (5.1,.1);
    \draw[ultra thick] (6,-.4) -- (6,-.2);
    \draw[ultra thick] (3,-.4) -- (3,-.2);

    \tikzstyle{vertex}=[circle, draw, text opacity = 1, inner sep=1pt]
    \tikzset{>=latex}
    \node[vertex](A) at (10.5, -1) {A};
    \node[vertex](T) at (11.5, .2) {T};
    \node[vertex](M) at (12.5, -1) {M};
    
    \begin{scope}[every path/.style={->}, every node/.style={inner sep=1pt}]
           \draw (A) -- (T);
           \draw (M) -- (T);
    \end{scope} 
    \end{tikzpicture}

    \caption{Score ranges and resulting Hasse diagram for the scenario in Example \ref{ex:gender}.}
    \label{fig:gender-example}
\end{figure}
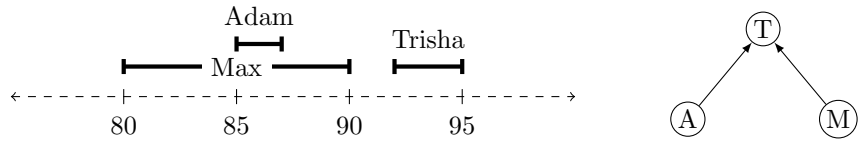
\end{example}

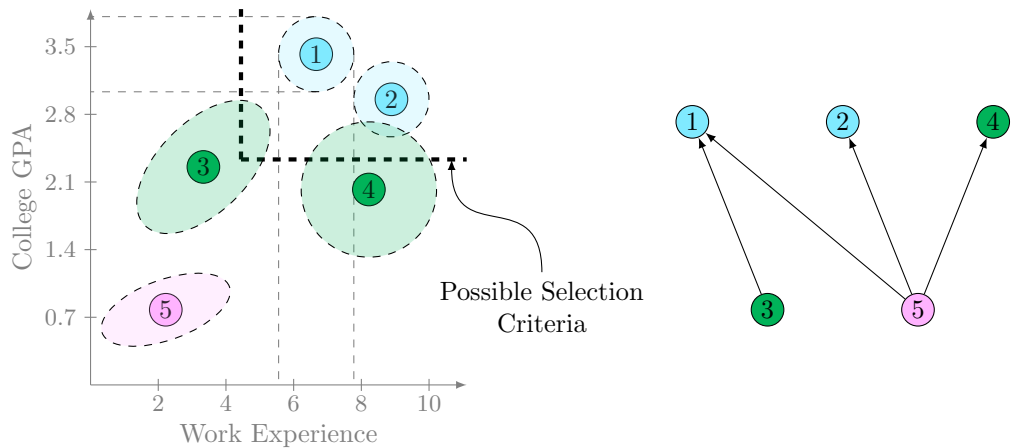
\begin{figure}
    \centering
    
    \begin{tikzpicture}[transform shape]
    
    \def\shift{8}
    \tikzset{>=latex}
    
    \draw[->,gray] (0,0) -- (5,0) node [midway, below=11pt, fill=white] {Work Experience};
    \draw[->,gray] (0,0) -- (0,5) node [midway, above=17pt, sloped, fill=white] {College GPA};
    
    \foreach[evaluate={\b=int(2*\a)},evaluate={\c=.9*\a}] \a in {1,...,5}{
    \draw[gray] (\c,-.08) -- (\c,.08);
    \node[gray] at (\c,-.25){{\small \b}};
    }
    
    \foreach[evaluate={\b=.8*\a},evaluate={\c=.9*\a},evaluate={\d=int(floor(.7*\a))},evaluate={\e=int(Mod(7*\a,10))}] \a in {1,...,5}{
    \draw[gray] (-.08,\c) -- (.08,\c);
    \node[gray] at (-.4,\c){{\small \d.\e}};
    }
    
    \draw[dashed,ultra thick] (2,3) -- (2,5);
    \draw[dashed,ultra thick] (2,3) -- (5,3);
    
    \definecolor{lightblue}{RGB}{128,234,255}
    \tikzstyle{vertex}=[circle, draw, fill=lightblue, inner sep=4pt]
    \node[vertex,inner sep = 1.5pt](b1) at (3, 4.4) {1};
    \node[vertex,inner sep = 1.5pt](rb1) at (8, 3.5) {1};
    
    \draw[thin,gray,dashed] (3,4.9) -- (0,4.9);
    % \node[gray] at (-.4,4.9) {4.0};
    \draw[thin,gray,dashed] (3,3.9) -- (0,3.9);
    % \node[gray] at (-.4,3.9) {3.5};
    
    \draw[thin,gray,dashed] (2.5,4.4) -- (2.5,0);
    % \node[gray] at (2.5,-.2) {4};
    \draw[thin,gray,dashed] (3.5,4.4) -- (3.5,0);
    % \node[gray] at (3.5,-.2) {5};
    
    \begin{scope}[shift={(3, 4.4)}]
      \draw[dashed, fill = lightblue, fill opacity = .2, rotate=45] (0,0) ellipse [x radius = .5 cm, y radius = .5 cm];
    \end{scope}
    
    \node[vertex,inner sep = 1.5pt](b2) at (4, 3.8) {2};
    \node[vertex,inner sep = 1.5pt](rb2) at (10, 3.5) {2};
    
    \begin{scope}[shift={(4, 3.8)}]
      \draw[dashed, fill = lightblue, fill opacity = .2, rotate=45] (0,0) ellipse [x radius = .5 cm, y radius = .5 cm];
    \end{scope}
    
    \definecolor{newgreen}{RGB}{0,179,89}
    \tikzstyle{vertex}=[circle, draw, fill=newgreen, inner sep=4pt]
    \node[vertex,inner sep = 1.5pt](g1) at (1.5, 2.9) {3};
    \node[vertex,inner sep = 1.5pt](rg1) at (9, 1) {3};
    
    \begin{scope}[shift={(1.5, 2.9)}]
      \draw[dashed, fill = newgreen, fill opacity = .2, rotate=45] (0,0) ellipse [x radius = 1.1 cm, y radius = .6 cm];
    \end{scope}
    
    \node[vertex,inner sep = 1.5pt](g2) at (3.7, 2.6) {4};
    \node[vertex,inner sep = 1.5pt](rg2) at (12, 3.5) {4};
    
    \begin{scope}[shift={(3.7, 2.6)}]
      \draw[dashed, fill = newgreen, fill opacity = .2, rotate=45] (0,0) ellipse [x radius = .9 cm, y radius = .9 cm];
    \end{scope}
    
    \definecolor{newpink}{RGB}{255,179,255}
    \tikzstyle{vertex}=[circle, draw, fill=newpink, inner sep=4pt]
    \node[vertex,inner sep = 1.5pt](p) at (1, 1) {5};
    \node[vertex,inner sep = 1.5pt](rp) at (11, 1) {5};
    
    % \draw[thin,gray,dashed] (2,1.5) -- (2,0);
    % \node[gray] at (2,-.2) {3};
    % \draw[thin,gray,dashed] (.5,.8) -- (.5,0);
    % \node[gray] at (3.5,-.2) {4};
    
    \begin{scope}[shift={(1, 1)}]
      \draw[dashed, fill = newpink, fill opacity = .2, rotate=20] (0,0) ellipse [x radius = .9 cm, y radius = .4 cm];
    \end{scope}
    
    \node at (6,1) {\begin{tabular}{c} Possible Selection \\ Criteria \end{tabular}};
    
    \draw [->] (6, 1.5) 
      .. controls ++(90:1.2) and ++(90:-1.2) .. (4.8,3);
    
    \draw[->] (rg1) -- (rb1);
    \draw[->] (rp) -- (rg2);
    \draw[->] (rp) -- (rb1);
    \draw[->] (rp) -- (rb2);
    
    % \begin{scope}[every path/.style={->,line width=.2pt}, every node/.style={inner sep=1pt}]
    % \path[thin,every node/.style={}](rg1) edge[bend left=20] (rb1);
    % % \path[thin,every node/.style={}](g1) edge[bend left=20] (b2);
    % \path[thin,every node/.style={}](rp) edge[bend right=20] (rg2);
    % \path[thin,every node/.style={}](rp) edge[bend right=10] (rb1);
    % \path[thin,every node/.style={}](rp) edge[bend right=10] (rb2);
    % \end{scope}

    \end{tikzpicture}
    
    \caption{A depiction of the work experience and college GPA of five candidates coming from three groups (differentiated by color). %Work experience is measured on the horizontal axis, and college GPA is measured on the vertical axis. 
    The thick dashed line represents a possible selection criteria which, in this case, imposes a threshold on each of the two attributes. Confidence regions are drawn around each data point to indicate, say, 95\% confidence in the inclusion of a candidate's true ability. Given these confidence regions, one can construct a partial ranking, as depicted by the Hasse diagram on the right. Arrows between candidates indicate ranking with certainty with respect to both attributes (e.g., Candidate 1 is ranked higher than Candidate 5 since the best work experience score ($\approx 4$) in the confidence region of Candidate 5 is worse than the worst work experience score ($\approx 5.5$) of Candidate 1, and similarly for College GPA. Note that there may be other reasonable ways of constructing partial rankings as well. %for every choice of scores within the candidates' respective confidence regions, Candidate 5 has better work experience and better GPA). %\sg{explain this is only one way to build the poset. there could be other ways. Add 0 on the left corner.} \js{Add numbers on axes and explain why 5 is worse than 1.}
    }
    \label{fig:2d-figure}
\end{figure}

\begin{example}
\label{ex:2d}
Suppose that three candidates are to be selected based on two attributes: work experience and college GPA. You have set cutoffs for each of these attributes and only wish to select candidates exceeding each cutoff. See Figure \ref{fig:2d-figure} for a depiction of the candidate pool, where each color represents a particular demographic group. Let the colored areas around each candidate node represent a ``confidence region;'' i.e., with some high degree of confidence, the candidate's latent ability lies in the drawn region. Note that we can infer partial rankings from these confidence regions in a similar way to Example \ref{ex:gender}: if the confidence region of candidate $A$ is strictly above and to the right of the confidence region of candidate $B$, then $A$ is ranked above $B$.

Using only raw scores, only the two blue candidates meet the cutoffs. However, taking confidence regions into account, we see that the two green candidates might meet the cutoffs as well. How, then, should one choose three candidates among the green and blue ones? One way to do so is to construct a partial ranking based on the confidence regions, as shown in Figure \ref{fig:2d-figure}. In this partial ranking, there are three candidates who are maximally ranked (i.e., are not ranked below any other candidates): the two blue candidates and the right-most green candidate. This observation could be one justification for selecting these candidates. 
\end{example}

% \begin{center}

% \end{center}

The process outlined in these examples (forming score ranges/regions for each candidate and inferring comparisons therefrom) can be applied quite generally, and allows for explicit treatment of bias in data. Data-driven techniques, such as estimating latent group-bias in a machine learning model, can be applied to generate these score ranges, which in turn induce a partial ranking. Such methods can be used to avoid penalizing applicants who come from underrepresented groups, who are more likely to face inaccurate evaluation via machine learning models. A recent paper by Emelianov et al. shows that groups with high error variances can receive worse treatment, even if the evaluations are unbiased for all candidates \cite{emelianov2020on}, pointing to the need for interventions like the poset approach that take uncertainty into account.

\paragraph{The poset approach in practice.} We end this section by providing a framework for using the poset approach in practice. While this framework does not encompass every possible use of the poset approach \cite{closingthegap}, it will describe the process from beginning to end and put the poset approach in broader context (see Figure \ref{fig:overview}).

\vspace{1.5mm}
\noindent 
\textbf{Step 1. Clean-up and Process Past Hiring Data}. To start, collect data from previous hiring cycles. This data might include scores derived from textual analysis of r\'esum\'es, test scores for job-related tasks (e.g., computer programming test scores), automated scores based on analysis of video interviews \cite{raghavan2020mitigating}, college GPAs, courses taken, years of work experience, job performance of those who were hired, and so on.

\vspace{1.5mm}
\noindent 
\textbf{Step 2. Quantify Uncertainty and Bias}. Use data analysis to quantify potential data biases. Clusterings, for example, can help determine if evaluations unfairly favor one group over another. Looking at the data along different demographics (e.g., based on race, gender, age) can point to potentially discriminatory decisions in the past. Use social science studies (e.g., \cite{steele}) that highlight the impact of social status on the considered metrics (e.g., standardized test scores). This will help highlight qualitative and quantitative reasons for disparities in the past hiring data.

\vspace{1.5mm}
\noindent 
\textbf{Step 3. Construct a Partial Order}. Trends identified in Step 2 can be used to construct a partial ranking of candidates. For example, score ranges can be constructed for each attribute of interest using a prediction model and estimates of its error variances. These ranges can take into account distributional differences across protected attributes, differing error variances due to training data imbalance,\footnote{This refers to the observation that a group which is underrepresented in training data often experiences large errors in a resulting prediction model. In the poset approach, these larger errors could translate to larger score ranges for the underrepresented group. Note that the groups in question could come from a clustering and need not be demographic groups.} observed inaccuracies in past predictions, and so on. Unsupervised methods such as clustering can be used without the specific knowledge about protected information, or this can be abstracted out by a third-party vendor to simply provide a hiring entity with the resultant estimate of uncertainties or the poset over the candidates. These approaches are discussed in more detail in Appendix \ref{app:constructing-the-poset}. The goal here is to account for uncertainties, inaccuracies, and biases in a direct and mathematically justified way, thereby paving the way to fairer decisions.

\vspace{1.5mm}
\noindent 
\textbf{Step 4. Adapt Selection Algorithms.} Once the partial ranking has been constructed, selections need to be made. Presumably, a hiring committee already has a screening process (automated or otherwise) which aligns with the goals of the employer. In order to implement the poset approach, this screening process must be adapted to take a partial ranking as input instead of numeric scores or a total ranking. Typically, this can be done by prioritizing maximality and randomizing wherever incomparabilities necessitate (see \cite{closingthegap} for an example of this in an online setting).

\vspace{1.5mm}
\noindent 
\textbf{Step 5. Auditing for Policy Compliance.} The entire hiring pipeline may be subject to auditing for compliance with anti-discrimination policy. It is prudent to document and be able to justify each decision made in the hiring process, particularly those pertaining to the four steps outlined above. For example, one should be able to explain how the partial ranking was constructed and be able to justify those decisions by pointing to data and relevant research. A deeper discussion of the legality of the poset approach (and algorithmic bias mitigation more generally) is in Section \ref{sec:law-4-8-21}.

\section{Discussion and Best Practices---Law, Mathematics, and Posets in Practice}
\label{sec:law-4-8-21}

We now return to the business cases with which we started and the tensions they present regarding diversity, equity, and legal interests. On the one hand, firms are seeking to address diversity regardless of a history of discrimination. On the other hand, when evidence of past or present practices creating barriers is found, companies addressing those practices are pursuing affirmative action plans. In general, a firm that does little to account for race, gender, and other protected classes may find it has created disparate impact; and yet, when that firm seeks to take protected classes into account, such steps may violate the ban on disparate treatment. 

We offer that in the unlikely case where a firm has no reason to believe that norms, traditions, or societal inequalities are negatively affecting the ability for members of a protected group to pass through the stages of the hiring pipeline, action may be possible under diversity interests but not required by law. At least two shifts point to increased diversity activity. First, many companies have made public commitment to large steps to address diversity in employment. Second, there is a new push for companies to disclose workforce diversity data, which has resulted in 82 of the top 100 companies doing so. The public imperative combined with the data supports companies taking the initiative to address workforce imbalances regardless of legal requirements to do so \cite{kempner_2021}. In contrast, as matter of affirmative action, a firm with evidence of discrimination seeking to address imbalances in its workforce should be able to take steps to do so. 
Such steps could involve, for instance, scoring applicants using a machine learning model and developing confidence intervals around scores using the poset approach. From a legal perspective, it is important to be able to \uline{support the legality of each action}, from the decision to address diversity to the decision to use protected class information, to each design choice in the algorithm, to each adjustment to future rounds of hiring. 

The beauty of the poset approach is that it is agnostic to the motivation,  diversity or addressing discrimination via affirmative action, behind a company's plan. To be clear, whether a purely diversity-driven plan is legal is an unsettled question and beyond the scope of this paper \cite{Estlund,bent2019algorithmic}. Nonetheless, because of the current drive to address inequity, we expect this question to arise in the near future and suggest that the poset approach would aid and support such efforts. Furthermore, because many announced diversity programs are likely backed by data about imbalances and unnecessary barriers to employment, such efforts will likely be seen as affirmative action plans under the law. Thus in this section we address the core question of how well the poset approach stands up to legal scrutiny as an allowed method to address affirmative actions plans. 

Given that efforts to modify evaluation mechanisms or selection algorithms can raise both disparate impact and disparate treatment issues, we now use a hypothetical employer perspective in line with Microsoft's and other companies' announced goals to suggest best practices. Insights are derived from a series of questions about how to identify workforce imbalances (Section \ref{subsec:diagnosis}) and how to address said imbalances (Section \ref{subsec:corrective-action}).

\subsection{Diagnosis}
\label{subsec:diagnosis}

\paragraph{\bf Q1:} {\it An employer is concerned that its workforce under-represents women and minorities. May they do anything to change their current hiring practices?}

\uline{Yes.} The purpose behind Title VII is ``[T]o achieve equality of employment opportunities,'' and Congress ``directed the thrust of the Act to the consequences of employment practices, not simply the motivation'' \cite{griggs1971duke2}. %Griggs v. Duke Power Co., 401 U.S. 424, 432 (1971).%  
\uline{That means ``unnecessary barriers to employment'' must fall, even if ``neutral on their face'' and ``neutral in terms of intent''} \cite{griggs1971duke3}. %Id. at 431. 
Federal courts have disallowed a host of hiring and promotion practices that ``operate[d] as `built in headwinds' for minority groups'' \cite{ricci5}. In addition, the Supreme Court has upheld the legality of employment plans to address discrimination without reference to its past practices or evidence of a possible violation of the law \cite{johnson_627}. 

To take action, an employer ``need[s] to point only to a ‘conspicuous ... imbalance in traditionally segregated job categories’ ” \cite{johnson_627}. Logically, this requirement implies that initial, proactive analysis identifying the imbalance problems can serve as justification for adjustments to hiring practices. As such, employers can and should use data science and analytics to identify the imbalance in their hiring pipeline that it seeks to address \cite{MacCarthy,Houser,Kim2017}. 

As one example, the employer can use human resources data to examine its employment practices. First, it can audit its current workforce and get fine-grained information about who works at the company and at what levels. Such an approach allows the company to look beyond simple questions such as ``Does it have an equal number of men and women in the workforce?'' Instead, the company can see the gender and minority makeup at different levels of employment such as upper management, upper-middle management, middle management, administration, hourly workers, contractors, and so on. Visualizing the data with pie-charts or heat maps will provide clear, vivid ways to see the current state of affairs. Second, after such a study, the company can see potential sources of issues. It may find that women and minorities rarely move beyond middle management, are rarely interviewed for promotion, or that screening to date has not selected, or under-selected, women and minorities for interviews to be potential employees. At a general level, these types of analyses support the case that there is something to fix. This gets us to the next step in the process.

\vspace{0.5cm}
\noindent 
{\bf Q2:} {\it If a company finds that women and minorities are rarely interviewed and further finds that screening to date has not selected, or under-selected, women and minorities for interviews to be potential employees, do these conditions support allowing an employer to use protected-class information to build or apply a bias-aware algorithm at the screening stage?} 

Identifying a problem with a screening process or a structural problem in the company’s workforce, reveals a clear “unnecessary barrier to employment” even if the algorithm is neutral on its face and in intent. For example, if men tend to be scored higher than women (e.g., as in Fig. \ref{fig:poset-not-quotas-unimodal}), then a facially neutral selection algorithm would disproportionately select men, even if true ability is similar across genders. In general, the identified, strong evidence of bias in current algorithmic sorting in the hiring process, including the screening stage, should constitute the sort of ``built in headwind[] for minority groups” that the law seeks to eliminate. With sufficient evidence of bias and systemic barriers to equality of employment opportunities, \uline{an employer can make a case for using bias-aware algorithms.} 

\subsection{Corrective action}
\label{subsec:corrective-action}

Voluntary action to comply with the goals of Title VII is not only allowed; it is favored \cite{johnson2}. 
Nonetheless, in some cases, trying to further the goals of Title VII to address discrimination raises the paradox where one approach looks like disparate impact and a corrective action looks like disparate treatment. What can a company actually do?\\

\noindent 
{\bf Q3:} {\it May an employer use protected-class information to increase diversity among interviewees?}

This question is complex as it entwines various parts of the process that need to be slowly unpacked. A recent case {\it Ricci v. DeStefano} \cite{ricci} illustrates some problems and provides guidance on allowed and prohibited actions.

\paragraph{Background.} In {\it Ricci v. DeStefano}, the City of New Haven had developed a test for firefighter promotion with the help and validation of experts. When administered, 77 people took the lieutenant exam: ``43 whites, 19 blacks, and 15 Hispanics. Of those, 34 candidates passed: 25 whites, 6 blacks, and 3 Hispanics.'' 41 people took the captain's exam: ``25 whites, 8 blacks, and 8 Hispanics. Of those, 22 candidates passed: 16 whites, 3 blacks, and 3 Hispanics.'' Despite the experts' opinions and validations of the test, the City rejected the results because the pass rate caused the city to believe it might be sued for disparate impact. The Supreme Court did not allow this after-the-fact change, because New Haven's actions relied on race, (the race of those who passed the test), to reject the results, and in that sense, New Haven engaged in disparate treatment. Thus, it may appear that an entity cannot account for and alter employment practices when there is evidence of potential disparate impact in the entity's practices, because such changes will necessarily constitute disparate treatment \cite{barocas2016big}. That is incorrect \cite{Kim}.

\paragraph{Analysis.} As the Supreme Court put it, not allowing an entity to account for race to avoid disparate impact liability ``if the employer knows its practice violates the disparate-impact provision,” is contrary to \uline{``Congress's intent that ``voluntary compliance'' be ``the preferred means of achieving the objectives of Title VII''} \cite{ricci2}. This rule, however, does not mean an entity can simply assert there has been a history of past discrimination and so a need to throw out a practice, because that might lead to ``an unyielding racial quota” \cite{ricci3}. As stated above, the entity has to show why the change is needed in light of the goals of Title VII. In addition, the timing of when an entity makes changes matters. 

The way the test was developed and administered by New Haven doomed the City's decision to reject the test's outcomes. New Haven began well by hiring experts to design a likely \emph{valid test}. The City spent \$100,000 on experts on designing the tests for fire departments \cite{ricci564}. The experts conducted interviews, went on ride-alongs, interviewed incumbents at the promotional level at issue, and designed ``job-analysis questionnaires and administered them to most of the incumbent battalion chiefs, captains, and lieutenants in the Department'' \cite{ricci564}. As the Supreme Court noted, ``At every stage of the job analyses, IOS [the company that developed the test], by deliberate choice, oversampled minority firefighters to ensure that the results---which IOS would use to develop the examinations---would not unintentionally favor white candidates'' \cite{ricci565}. Once the test was approved, New Haven set a 3-month study period and gave candidates a study guide including the ``source material for the questions, including the specific chapters from which the questions were taken'' \cite{ricci565}. Nonetheless, after the tests were given, the results indicated disparate impact \cite{ricci567}. 

The city's ex-post actions were the problem. The Court rejected ``invalidating the test results'' after the fact without ``a strong basis in evidence of an impermissible disparate impact'' \cite{ricci4}. The ex-post rejection of the results created ``visible victims''---that is, those who studied for the test, passed, and whose hard work was discarded \cite{Primus}. After the city gave the test, it needed strong evidence that the test would be invalidated if the city were sued for disparate impact and lose, because otherwise those who had passed would be harmed. The Court did not see such evidence and so did not allow the city to reject the results.

\paragraph{Answer to Q3.} Designing a screening system is quite different than what happened in {\it Ricci}. {\it Ricci} was about a later stage of employment (i.e., promotions), and it involved a test for which many test-takers had prepared, including spending money on test preparation aid. \uline{The advantage of building a screening system is that the actions are ex-ante, and the system is not a test for which someone can prepare} \cite{bent2019algorithmic}. Unlike in {\it Ricci}, where applicants were seen as having an expectation that a potentially valid test for which they could study be accepted, designing and using a screening algorithm occurs at an earlier stage of the hiring process where no hiring or promotion decision is made. Thus in designing a screening algorithm, one might observe selections over time and change the parameters to create a more representative sample of qualified candidates, including making adjustments during the ``training'' of the algorithm. These steps are analogous to the design steps---such as making overt choices and oversampling at every stage to ensure that the test did ``not unintentionally favor white candidates''---taken by New Haven and of which the Supreme Court wrote with approval \cite{ricci565}. In other words, designing and vetting a screening system to ensure that the results are not having discriminatory outcomes should be legal. 

\uline{Recall that one of the goals of Title VII is to reduce, if not eliminate, ``unnecessary barriers to employment.''} The {\it Ricci} Court did not ``question an employer's affirmative efforts to ensure that all groups have a fair opportunity'' at a given stage of the hiring process. An employer is allowed to examine ``how to design$\ldots$[a] practice in order to provide a fair opportunity for all individuals, regardless of their race'' before deploying it \cite{ricci4}. Designing a screening algorithm is by its nature an ex-ante event for which a candidate cannot prepare in the way one might for a test.

In short, if Question 2's requirement is met, an employer should be able to develop a bias-aware algorithm to avoid disparate impact. Of course, we still need to address the validity of the new practice and what is allowed in its design, which brings us to the next question, which we partially answer through the lens of the poset approach.

\vspace{0.5cm}
\noindent 
{\bf Q4:} {\it What is allowed in the design of a bias-aware algorithm? Can it be designed to improve the yield of whom to interview?}

This is one of the grand challenges in this area. Let us focus our attention to the proposed poset approach, and draw arguments from the Supreme Court's decision in {\it Johnson}. The key to using a bias-aware algorithm such as the poset approach of Salem and Gupta is to establish the facts and evidence of a need to address bias (or more generally, inconsistencies in the data) as set forth above, and then to build a plan that \uline{assesses individuals rather than setting up a purely number-driven process with quotas for each category} \cite{johnson2}. If a plan is ``blind hiring,'' that is, dictates hiring ``solely by reference to statistics'' or ``by reflexive adherence to a numerical standard,'' the plan is not likely to be allowed \cite{johnson3}. %Johnson 480 US at 636-637.%
\uline{But, if a plan takes ``numerous factors$\ldots$into account in making hiring decisions, including specifically the qualifications of [all] applicants for particular jobs,'' the plan may take a protected class into account as part of the overall evaluation} \cite{johnson2}. %(Johnson 480 US at 631).
In that sense, the protected class status “may be deemed a ‘plus’ in a particular applicant’s file, yet it does not insulate the individual from comparison with all other candidates for the available seats'' \cite{johnson638,bakke}.

Comparison does not require pure, numeric ranking; indeed, that might tip into the sort of ``blind hiring'' that is disfavored. As the Sixth Circuit stated, the ``practice of rank-order hiring from a single list grouping together males and females was impermissible under Title VII because the City could not establish that higher scores on the test meant better job performance.'' \cite{brunetcolumbus}. The Second Circuit has explained that evaluations should be sufficiently correlated with job performance to induce a rank ordering, where the quantification of ``sufficiently correlated'' may depend on the extent of adverse impact of the evaluation metric \cite{guardians}. The Sixth Circuit additionally asserted that a certain cognitive ability test could not be used as the sole basis for a rank-ordering despite being predictive of job performance, since the test failed to measure certain qualities of interest. Rank orderings based on evaluations should therefore not be thought of as implicit to a screening practice, but instead as a design choice which must be justified \cite{brunetcolumbus}.

Discretion in comparison of candidates is allowed when it is part of the overall, individual assessment. For example, in {\it Johnson v. Transportation Agency of Santa Clara County}, two candidates were deemed well-qualified based on a range of metrics, such as experience, background, and test scores taken together. But each candidate had differences within a given metric. One had more clerical work and more road maintenance work; the other had more experience at a specific part of the business. As for test scores, the man scored 75 on the interview portion of the assessment and the woman scored 73. The employer had set 70 as the minimum threshold for the interview and seven applicants crossed the 70 mark. The range of acceptable scores was 70 to 80 \cite{johnson623}. The woman was given the promotion over the man who had the higher score. \uline{Because the scores were within the range of acceptable scores and the final hiring manager looked at a set of metrics with gender as ``but one of numerous factors he took into account in arriving at his decision,'' the plan's incorporation of bias-awareness, here gender, was allowed} \cite{johnson638}.

Other cases also acknowledge the need for an approach beyond using an absolute score or ranking. Given problems with rank-ordering, the Second Circuit of Appeals has allowed a rather coarse approach where an employer may ``acknowledge his inability to justify rank-ordering and resort to random selection from within either the entire group that achieves a properly determined passing score, or some segment of the passing group shown to be appropriate'' \cite{kirkland}. Courts have also indicated an acceptance for more nuanced methods. For example, the act of ``banding,'' or considering score ranges instead of singular scores, has been accepted to account for inaccuracies in evaluation. \cite{bradley, boston}. Although these cases consider banding in a quite limited sense in that scores ranges are centered on original scores and are of uniform length, they support that one might relax the assumption of an absolute ranking of candidates. 

In language that tracks the poset approach, the Second Circuit has also acknowledged “that small differences between the scores of candidates indicate very little about the candidates’ relative merit and fitness'' \cite{kirkland}. Thus the court embraced an approach that assessed ``a statistical computation of the likely error of measurement inherent'' in its exam. The employer then used that measurement to set up zones of candidates clustered by test scores within that error measurement. That practice was seen as a good solution to ``insur[e] compliance'' with Title VII.  The Second Circuit explained, ``by creating a more valid method to assess the significance of test scores, [the approach] eliminated the central cause of the adverse impact, i.e., the rank-ordering system, while assuring appointments on the basis of merit.'' As such, if one is able to use protected information (as in \emph{Johnson}, or in the context of a valid affirmative action plan \cite{bent2019algorithmic}), then the banding cases provide guideposts for adopting the poset approach as described in Section \ref{sec:new-directions}.

\paragraph{Answer to Q4.} \uline{1. An algorithmic approach should be allowed.} A takeaway from {\it Johnson} and the cases on banding and rank-ordering is that a precise numerical score is not necessarily indicative of an applicant's potential, and courts welcome approaches that better compare candidates. Thus, score ranges can be used as part of an applicant-screening procedure. This supports the use of score ranges to account for uncertainties in evaluations, as outlined in Section 3. 

Further, note that incorporating the poset model of bias is not the same thing as normalizing distributions of scores across groups. When we normalize scores across groups, we are essentially transforming all scores so that group-specific distributions look similar, and this process results in a full ranking of applicants. In contrast, the poset approach intentionally does not reduce each applicant to a number and allows for incomparabilities between applicants. This allows for a more individual treatment of candidates, where uncertainty in rankings can be acknowledged. The result is that applicants are assessed as individuals, potentially in a more mathematically sound way. 

\uline{2. There are rules about when bias-aware algorithms can be used.} Recall that the stage at which an entity uses bias-aware algorithms matters. In the promotion context of {\it Johnson}, the Court gave a further reason the plan was allowed. Unlike {\it Ricci}, where applicants were seen as having an expectation that a potentially valid test for which they could study be accepted, there was ``no absolute entitlement'' to the position at issue in {\it Johnson}. The entity had seven qualified and eligible applicants, and choosing one over the other ``unsettled no legitimate, firmly rooted expectation'' of any of the candidates. By extension, a bias-aware applicant-screening plan that used a protected class as part of an overall assessment then had all selected applicants compete on the same metrics should be allowed under the law. 

\uline{3. There are legal rules on the goals of any hiring plan.}
The law respects plans that seek to remedy an imbalance and that do not set aside positions for a given group while also conducting annual reviews of goals as it fashions future rounds of hiring and promotion \cite{johnson640}.  One may work ``to attain a balanced work force, not to maintain one'' \cite{johnson639}.

The {\it Johnson} Court also noted with approval that ``the Plan sought annually to develop even more refined measures of the under-representation in each job category that required attention'' \cite{johnson635}. This idea of not maintaining a balanced workforce reflects the idea that an entity cannot use a plan that sets up quotas to maintain balance based purely on class statuses. By extension, suppose balance is achieved in a company through bias-aware methods, and they notice this by continuous monitoring of their hiring practices (in a sense, returning to Question 1). The company may then have to stop using bias-aware methods, even if demographic imbalance persists in the general workforce for that line of work. 

\uline{4. The poset approach does not impose quotas.}
In contrast to methods described in some recent work (e.g., \cite{emelianov2020on}), using score ranges (i.e., the poset approach) instead of raw scores does not set up a quota system.\footnote{There is some debate on whether quotas are allowed at a screening stage. One reason to not use quota-based approaches at screening stages is to avoid setting up a pre-destined pool of candidates who will be later rejected by the system. The poset approach accounts for uncertainties in candidate evaluations, and makes selections based on the possibility of a candidate being qualified, and this process can be irrespective of demographic features. This is attractive from a legal perspective at any stage in the hiring pipeline.} When using the poset approach, selection rates may be influenced by protected information (e.g., when accounting for observed, group-specific biases), but such protected information is not necessarily a determining factor in selection decisions. For example, the poset approach could result in a set of candidates which is less demographically proportional than what raw scores might produce dependent on the data and ascertained uncertainty in the data (see, e.g., Figures \ref{fig:poset-not-quotas-separated}-\ref{fig:poset-not-quotas-bimodal-nonmonotonic} in Appendix \ref{app:poset-not-quotas}) or could result in more demographically proportional selections (see, e.g.,  Figures \ref{fig:poset-not-quotas-unimodal}-\ref{fig:poset-not-quotas-bimodal}). It simply accounts for the uncertainty in the candidate evaluations.

\section{Conclusion}

We summarized recent work in the context of hiring, with a focus on screening algorithms. We highlighted the seeming paradox of mathematics, law and practice that a company might observe workforce imbalance due to its past practices, but the solutions to correct for this imbalance are either at a contradiction with mathematics or anti-discrimination law. The new poset-based approach (analyzed with a streaming model in \cite{closingthegap}) provides a framework for incorporating uncertainties in rankings into a candidate-screening practice which allows, for example, hiring committees to base decision on confidence intervals of ability scores. This approach can potentially be legally justified based on past disparate impact and can be adjusted over time as the data grows and hiring goals evolve; and thus can help avoid having a static plan as the law requires.

No approach, however, is a fix-all solution. The poset approach cannot discount for undetectable errors undetectable, or modeling errors due to missing data. The ranges of the intervals impact the quality of selections. Further, two different mathematical approaches could be used to define score ranges for candidates and result in different sets of selected candidates. A legal dispute may require addressing which one of these approaches is more \uline{valid}. Further, there is an ``are we there yet?'' issue built into the Supreme Court's rulings. That is, it may be unclear at which point a workforce becomes ``balanced'' and the current plan must be replaced. Although the poset approach is adaptive, detecting where there is no longer any impact of societal biases in the data is non-trivial and we leave this as an open question. 

For any intervention in an existing framework, one has to consider if the intervention is serving those for whom it is designed \cite{goel2018}. Partially ordered sets that are interval-based might create an impression that certain underrepresented minorities carry high uncertainty in their potentials and as a result, lead a risk-averse hiring committee to reject those candidates. On the contrary, the poset approach can highlight {\it missed opportunities} in representation in the hiring pipeline. Taking uncertainties into account can 
expand and improve the talent pool to include candidates who are qualified and would have been competitive had there been no bias in the data. Thus, we believe that the analysis presented here can pave the way forward for hiring qualified candidates in a fair way in the evolving legal landscape.

\section*{Acknowledgments}

The authors thank Jason Bent, Justin Biddle, Kimberly Houser, Pauline Kim, Orly Lobel, and the participants of the Data, Law, and Ethics Virtual Conference hosted by the University of Indiana, Kelly School of Business, as well as the participants of the Privacy Law Scholars Conference 2021 for their helpful comments on an earlier draft of this work.

%%
%% The next two lines define the bibliography style to be used, and
%% the bibliography file.
\bibliographystyle{abbrv}
\bibliography{refs}

\begin{thebibliography}{100}

\bibitem{griggs1971duke3}
{\em \textup{Griggs v. Duke Power Co., 401 U.S. 424, 431}}.
\newblock 1971.

\bibitem{griggs1971duke2}
{\em \textup{Griggs v. Duke Power Co., 401 U.S. 424, 432}}.
\newblock 1971.

\bibitem{bakke}
{\em \textup{Regents of University of California v. Bakke, 438 U.S.
  \texttt{265, 317}}}.
\newblock 1978.

\bibitem{eeoc29cfr}
{\em \textup{EEOC Guidelines on Affirmative Action, 29 C.F.R. § 1608.1(c)}}.
\newblock 1979.

\bibitem{guardians}
{\em \textup{Guardians Ass'n of New York City v. Civil Serv, 630 F.2d 79 (2d
  Cir.)}}.
\newblock 1980.

\bibitem{kirkland}
{\em \textup{Kirkland v. N.Y. State Dep’t of Correctional Serv., 711 F.2d
  1117, 1133 (2d Cir.)}}.
\newblock 1983.

\bibitem{johnson623}
{\em \textup{Johnson v. Transportation Agency, Santa Clara Cty., 480 U.S.
  \texttt{616, 623-624}}}.
\newblock 1987.

\bibitem{johnson_627}
{\em \textup{Johnson v. Transportation Agency, Santa Clara Cty., 480 U.S.
  \texttt{616, 627}}}.
\newblock 1987.

\bibitem{johnson2}
{\em \textup{Johnson v. Transportation Agency, Santa Clara Cty., 480 U.S.
  \texttt{616, 631}}}.
\newblock 1987.

\bibitem{johnson635}
{\em \textup{Johnson v. Transportation Agency, Santa Clara Cty., 480 U.S.
  \texttt{616, 635}}}.
\newblock 1987.

\bibitem{johnson3}
{\em \textup{Johnson v. Transportation Agency, Santa Clara Cty., 480 U.S.
  \texttt{616, 636-637}}}.
\newblock 1987.

\bibitem{johnson638}
{\em \textup{Johnson v. Transportation Agency, Santa Clara Cty., 480 U.S.
  \texttt{616, 638}}}.
\newblock 1987.

\bibitem{johnson639}
{\em \textup{Johnson v. Transportation Agency, Santa Clara Cty., 480 U.S.
  \texttt{616, 639}}}.
\newblock 1987.

\bibitem{johnson640}
{\em \textup{Johnson v. Transportation Agency, Santa Clara Cty., 480 U.S.
  \texttt{616, 640-641}}}.
\newblock 1987.

\bibitem{brunetcolumbus}
{\em \textup{Brunet v. City of Columbus, Ohio, 58 F.3d 251, 255 (6th Cir.)}}.
\newblock 1995.

\bibitem{boston}
{\em \textup{Boston Police Superior Officers Fed’n v. City of Boston, 147
  F.3d 13}}.
\newblock 1998.

\bibitem{bradley}
{\em \textup{Bradley v. City of Lynn, 403 F. Supp. 2d 161}}.
\newblock 2005.

\bibitem{eeoc}
{\em \textup{Title VII, 29 CFR Parts 1600, 1607, 1608}}.
\newblock 2006.

\bibitem{ricci5}
{\em \textup{Ricci v. DeStefano, 557 U.S. 557, 632}}.
\newblock 2009.

\bibitem{ricci}
{\em \textup{Ricci v. DeStefano, 557 U.S. \texttt{557}}}.
\newblock 2009.

\bibitem{ricci564}
{\em \textup{Ricci v. DeStefano, 557 U.S. \texttt{557, 564}}}.
\newblock 2009.

\bibitem{ricci565}
{\em \textup{Ricci v. DeStefano, 557 U.S. \texttt{557, 565}}}.
\newblock 2009.

\bibitem{ricci567}
{\em \textup{Ricci v. DeStefano, 557 U.S. \texttt{557, 567}}}.
\newblock 2009.

\bibitem{ricci2}
{\em \textup{Ricci v. DeStefano, 557 U.S. \texttt{557, 580-581}}}.
\newblock 2009.

\bibitem{ricci3}
{\em \textup{Ricci v. DeStefano, 557 U.S. \texttt{557, 583}}}.
\newblock 2009.

\bibitem{ricci4}
{\em \textup{Ricci v. DeStefano, 557 U.S. \texttt{557, 585}}}.
\newblock 2009.

\bibitem{bostock}
{\em \textup{Bostock v. Clayton County, 590 U.S. \texttt{\_\_\_}}}.
\newblock 2020.

\bibitem{accord-42}
{\em \textup{§ 2000e-2(k)(1)(A).}}
\newblock 42 U.S.C.

\bibitem{mckinsey}
Mckinsey's online application faqs: Careers, Accessed January 17, 2022.

\bibitem{angwin2017dozens}
J.~Angwin, N.~Scheiber, and A.~Tobin.
\newblock Dozens of companies are using facebook to exclude older workers from
  job ads.
\newblock {\em ProPublica, December}, 2017.

\bibitem{barocas2014data}
S.~Barocas.
\newblock Data mining and the discourse on discrimination.
\newblock In {\em Data Ethics Workshop, Conference on Knowledge Discovery and
  Data Mining}, pages 1--4, 2014.

\bibitem{barocas-hardt-narayanan}
S.~Barocas, M.~Hardt, and A.~Narayanan.
\newblock {\em Fairness and Machine Learning}.
\newblock fairmlbook.org, 2019.
\newblock \url{http://www.fairmlbook.org}.

\bibitem{barocas2016big}
S.~Barocas and A.~D. Selbst.
\newblock Big data's disparate impact.
\newblock {\em Calif. L. Rev.}, 104:671, 2016.

\bibitem{bass_2020}
D.~Bass and J.~Eidelson.
\newblock Microsoft plan to add black executives draws u.s. labor inquiry.
\newblock {\em Seattle Times}, October 6, 2020.

\bibitem{batastini2017bias}
A.~B. Batastini, A.~D. Bola{\~n}os, R.~D. Morgan, and S.~M. Mitchell.
\newblock Bias in hiring applicants with mental illness and criminal justice
  involvement: A follow-up study with employers.
\newblock {\em Criminal Justice and Behavior}, 44(6):777--795, 2017.

\bibitem{bent2019algorithmic}
J.~R. Bent.
\newblock Is algorithmic affirmative action legal.
\newblock {\em Geo. LJ}, 108:803, 2019.

\bibitem{blum2020recovering}
A.~Blum and K.~Stangl.
\newblock Recovering from biased data: Can fairness constraints improve
  accuracy?
\newblock In {\em Symposium on Foundations of Responsible Computing (FORC)},
  volume~1, 2020.

\bibitem{bogen2018help}
M.~Bogen and A.~Rieke.
\newblock Help wanted: An examination of hiring algorithms, equity, and bias.
\newblock 2018.

\bibitem{caliskan2017semantics}
A.~Caliskan, J.~J. Bryson, and A.~Narayanan.
\newblock Semantics derived automatically from language corpora contain
  human-like biases.
\newblock {\em Science}, 356(6334):183--186, 2017.

\bibitem{celis2020interventions}
L.~E. Celis, A.~Mehrotra, and N.~K. Vishnoi.
\newblock Interventions for ranking in the presence of implicit bias.
\newblock In {\em Proceedings of the 2020 Conference on Fairness,
  Accountability, and Transparency}, pages 369--380, 2020.

\bibitem{christensen1996analysis}
R.~Christensen.
\newblock {\em Analysis of variance, design, and regression: applied
  statistical methods}.
\newblock Page 173. CRC Press, 1996.

\bibitem{goel2018}
S.~Corbett-Davies and S.~Goel.
\newblock The measure and mismeasure of fairness: A critical review of fair
  machine learning.
\newblock {\em arXiv preprint arXiv:1808.00023}, 2018.

\bibitem{corbett2017algorithmic}
S.~Corbett-Davies, E.~Pierson, A.~Feller, S.~Goel, and A.~Huq.
\newblock Algorithmic decision making and the cost of fairness.
\newblock In {\em Proceedings of the 23rd ACM SIGKDD International Conference
  on Knowledge Discovery and Data Mining}, pages 797--806. ACM, 2017.

\bibitem{glassdoor}
V.~Das~Swain, K.~Saha, M.~D. Reddy, H.~Rajvanshy, G.~D. Abowd, and
  M.~De~Choudhury.
\newblock Modeling organizational culture with workplace experiences shared on
  glassdoor.
\newblock In {\em Proceedings of the 2020 CHI Conference on Human Factors in
  Computing Systems}, pages 1--15, 2020.

\bibitem{dastin_2018}
J.~Dastin.
\newblock Amazon scraps secret ai recruiting tool that showed bias against
  women.
\newblock {\em Reuters}, Oct 2018.

\bibitem{de2019bias}
M.~De-Arteaga, A.~Romanov, H.~Wallach, J.~Chayes, C.~Borgs, A.~Chouldechova,
  S.~Geyik, K.~Kenthapadi, and A.~T. Kalai.
\newblock Bias in bios: A case study of semantic representation bias in a
  high-stakes setting.
\newblock In {\em Proceedings of the Conference on Fairness, Accountability,
  and Transparency}, pages 120--128, 2019.

\bibitem{desai2017trust}
D.~R. Desai and J.~A. Kroll.
\newblock Trust but verify: A guide to algorithms and the law.
\newblock {\em Harv. JL \& Tech.}, 31:1, 2017.

\bibitem{dewan_2014}
S.~Dewan.
\newblock How businesses use your sats.
\newblock {\em New York Times}, Mar 2014.

\bibitem{dixon2018measuring}
L.~Dixon, J.~Li, J.~Sorensen, N.~Thain, and L.~Vasserman.
\newblock Measuring and mitigating unintended bias in text classification.
\newblock In {\em Proceedings of the 2018 AAAI/ACM Conference on AI, Ethics,
  and Society}, pages 67--73, 2018.

\bibitem{SAT-Intersection}
E.~Dixon-Roman, H.~Everson, and J.~Mcardle.
\newblock {Race, Poverty and SAT Scores: Modeling the Influences of Family
  Income on Black and White High School Students' SAT Performance}.
\newblock {\em Teachers College Record}, 115, 05 2013.

\bibitem{Domingos}
P.~Domingos.
\newblock {\em The Master Algorithm: How the Quest for the Ultimate Learning
  Machine Will Remake Our World}.
\newblock Basic Books, New York, NY, 1st. edition, 2015.

\bibitem{duffy_2020_3}
C.~Duffy.
\newblock Adidas says at least 30\% of new us positions will be filled by black
  or latinx people.
\newblock {\em CNN Business}, June 9, 2020.

\bibitem{duffy_2020}
C.~Duffy.
\newblock In the face of a cultural reckoning, it turns out massive
  corporations can move fast and fix things.
\newblock {\em CNN Business}, June 21, 2020.

\bibitem{duffy_2020_2}
C.~Duffy.
\newblock Plans at microsoft and wells fargo to increase black leadership are
  under scrutiny from the labor dept.
\newblock {\em CNN Business}, October 7, 2020.

\bibitem{dwork}
C.~Dwork, M.~Hardt, T.~Pitassi, O.~Reingold, and R.~Zemel.
\newblock Fairness through awareness.
\newblock In {\em Proceedings of the 3rd innovations in theoretical computer
  science conference}, pages 214--226, 2012.

\bibitem{edelman2017racial}
B.~Edelman, M.~Luca, and D.~Svirsky.
\newblock Racial discrimination in the sharing economy: Evidence from a field
  experiment.
\newblock {\em American Economic Journal: Applied Economics}, 9(2):1--22, 2017.

\bibitem{emelianov2020on}
V.~Emelianov, N.~Gast, K.~P. Gummadi, and P.~Loiseau.
\newblock On fair selection in the presence of implicit variance.
\newblock In {\em Proceedings of the 2020 ACM Conference on Economics and
  Computation}, 2020.

\bibitem{Estlund}
C.~L. Estlund.
\newblock Putting grutter to work: Diversity, integration, and affirmative
  action.
\newblock {\em Berkeley J. of Labor and Employment}, 26:1, 2005.

\bibitem{faenza2020impact}
Y.~Faenza, S.~Gupta, and X.~Zhang.
\newblock Impact of bias on school admissions and targeted interventions, 2020.

\bibitem{fischer2007effects}
M.~J. Fischer and D.~S. Massey.
\newblock The effects of affirmative action in higher education.
\newblock {\em Social Science Research}, 36(2):531--549, 2007.

\bibitem{friedler2019comparative}
S.~A. Friedler, C.~Scheidegger, S.~Venkatasubramanian, S.~Choudhary, E.~P.
  Hamilton, and D.~Roth.
\newblock A comparative study of fairness-enhancing interventions in machine
  learning.
\newblock In {\em Proceedings of the conference on fairness, accountability,
  and transparency}, pages 329--338, 2019.

\bibitem{ftc-big-data-report}
{FTC Report}.
\newblock {\em Big data: a tool for inclusion or exclusion?}
\newblock Federal Trade Commission, January 2016.

\bibitem{goodman2018why}
R.~Goodman.
\newblock Why amazon's automated hiring tool discriminated against women, 2018.
\newblock Published Oct. 12, 2018. Last accessed Jun. 6, 2019.

\bibitem{griswold_2014}
A.~Griswold.
\newblock Why major companies like amazon ask job candidates for their sat
  scores.
\newblock {\em Business Insider}, Mar 2014.

\bibitem{hanks2009technology}
C.~Hanks.
\newblock {\em Technology and values: Essential readings}, page~7.
\newblock John Wiley \& Sons, 2009.

\bibitem{Hanna_and_Linden}
R.~N. Hanna and L.~L. Linden.
\newblock Discrimination in grading.
\newblock {\em American Economic Journal: Economic Policy}, 4:146--68, 02 2012.

\bibitem{hannak2017bias}
A.~Hann{\'a}k, C.~Wagner, D.~Garcia, A.~Mislove, M.~Strohmaier, and C.~Wilson.
\newblock Bias in online freelance marketplaces: Evidence from taskrabbit and
  fiverr.
\newblock In {\em Proceedings of the 2017 ACM Conference on Computer Supported
  Cooperative Work and Social Computing}, pages 1914--1933, 2017.

\bibitem{hardt2016equality}
M.~Hardt, E.~Price, and N.~Srebro.
\newblock Equality of opportunity in supervised learning.
\newblock In {\em Advances in neural information processing systems}, pages
  3315--3323, 2016.

\bibitem{heilman1997affirmative}
M.~E. Heilman, C.~J. Block, and P.~Stathatos.
\newblock The affirmative action stigma of incompetence: Effects of performance
  information ambiguity.
\newblock {\em Academy of Management Journal}, 40(3):603--625, 1997.

\bibitem{Houser}
K.~A. Houser.
\newblock Can ai solve the diversity problem in the tech industry? mitigating
  noise and bias in employment decision-making.
\newblock {\em Stanford Tech. L. Rev.}, 22:290, 2019.

\bibitem{jobsearch-rankings}
Jobscan.
\newblock Applicant tracking systems, Accessed Sept. 11, 2020.
\newblock Available at \url{https://www.jobscan.co/applicant-tracking-systems}.

\bibitem{kamiran2010discrimination}
F.~Kamiran, T.~Calders, and M.~Pechenizkiy.
\newblock Discrimination aware decision tree learning.
\newblock In {\em 2010 IEEE International Conference on Data Mining}, pages
  869--874. IEEE, 2010.

\bibitem{kamishima2012fairness}
T.~Kamishima, S.~Akaho, H.~Asoh, and J.~Sakuma.
\newblock Fairness-aware classifier with prejudice remover regularizer.
\newblock In {\em Joint European Conference on Machine Learning and Knowledge
  Discovery in Databases}, pages 35--50. Springer, 2012.

\bibitem{Kearns}
M.~Kearns and A.~Roth.
\newblock {\em The Ethical Algorithm: The Science of Socially Aware Algorithm
  Design}.
\newblock oxford University Press, New York, NY, 1st. edition, 2020.

\bibitem{kempner_2021}
M.~Kempner.
\newblock Georgia’s big businesses reveal staff — and management —
  diversit.
\newblock {\em The Atlanta Constitution Journal}, October 8, 2021.

\bibitem{Kim2017}
P.~Kim.
\newblock Auditing algorithms for discrimination.
\newblock {\em U. Pa. L. Rev. Online}, 166:189, 2017.

\bibitem{Kim}
P.~Kim.
\newblock Data-driven discrimination at work.
\newblock {\em Wm. \& Mary L. Rev.}, 58:8657, 2017.

\bibitem{kim2020manipulating}
P.~T. Kim.
\newblock Manipulating opportunity.
\newblock {\em Va. L. Rev.}, 106:867, 2020.

\bibitem{kleinberg2019simplicity}
J.~Kleinberg and S.~Mullainathan.
\newblock Simplicity creates inequity: implications for fairness, stereotypes,
  and interpretability.
\newblock In {\em Proceedings of the 2019 ACM Conference on Economics and
  Computation}, pages 807--808, 2019.

\bibitem{kr18}
J.~M. Kleinberg and M.~Raghavan.
\newblock Selection problems in the presence of implicit bias.
\newblock In {\em 9th Innovations in Theoretical Computer Science Conference,
  {ITCS} 2018, January 11-14, 2018, Cambridge, MA, {USA}}, pages 33:1--33:17,
  2018.

\bibitem{lum2016predict}
K.~Lum and W.~Isaac.
\newblock To predict and serve?
\newblock {\em Significance}, 13(5):14--19, 2016.

\bibitem{MacCarthy}
M.~MacCarthy.
\newblock Standards of fairness for disparate impact assessment of big data
  algorithms.
\newblock {\em Cumberland L. Rev.}, 48:102, 2017.

\bibitem{March}
J.~March.
\newblock Exploration and exploitation in organizational learning.
\newblock {\em Organizational Science}, 2:71, 1989.

\bibitem{miao2013properties}
W.~Miao and J.~L. Gastwirth.
\newblock Properties of statistical tests appropriate for the analysis of data
  in disparate impact cases.
\newblock {\em Law, Probability and Risk}, 12(1):37--61, 2013.

\bibitem{mohr2014women}
T.~S. Mohr.
\newblock Why women don’t apply for jobs unless they’re 100\% qualified.
\newblock {\em Harvard Business Review}, 25, 2014.

\bibitem{yale}
C.~Moss-Racusin, D.~, V.~Brescoll, M.~Graham, and J.~Handelsman.
\newblock Science faculty's subtle gender biases favor male students.
\newblock {\em Proceedings of the National Academy of Sciences},
  109:16474--16479, 09 2012.

\bibitem{U.S._Department_of_Labor}
U.~D. of~Labor Office~of Federal Contract Compliance~Programs.
\newblock U.s. department of labor and microsoft corp. enter agreement to
  resolve alleged hiring discrimination affecting 1,229 applicants in four
  states.
\newblock {\em CNN Business}, September 18, 2020.

\bibitem{Primus}
R.~Primus.
\newblock The future of disparate impact.
\newblock {\em Mich. L. Rev.}, 108:1341, 2010.

\bibitem{raghavan2020mitigating}
M.~Raghavan, S.~Barocas, J.~Kleinberg, and K.~Levy.
\newblock Mitigating bias in algorithmic hiring: Evaluating claims and
  practices.
\newblock In {\em Proceedings of the 2020 Conference on Fairness,
  Accountability, and Transparency}, pages 469--481, 2020.

\bibitem{closingthegap}
J.~Salem and S.~Gupta.
\newblock Closing the gap: Group-aware parallelization for online selection of
  candidates with biased evaluations.
\newblock In {\em {International Conference on Web and Internet Economics
  (WINE)}}. Springer, 2020. Under major revision at Management Science, 2021.

\bibitem{10.1145/3351095.3372849}
J.~S\'{a}nchez-Monedero, L.~Dencik, and L.~Edwards.
\newblock What does it mean to “solve” the problem of discrimination in
  hiring? social, technical and legal perspectives from the uk on automated
  hiring systems.
\newblock In {\em Proceedings of the 2020 Conference on Fairness,
  Accountability, and Transparency}, FAT* ’20, page 458–468, New York, NY,
  USA, 2020. Association for Computing Machinery.

\bibitem{schlachter2019employee}
S.~D. Schlachter and J.~R. Pieper.
\newblock Employee referral hiring in organizations: An integrative conceptual
  review, model, and agenda for future research.
\newblock {\em Journal of Applied Psychology}, 2019.

\bibitem{schwartz2019untold}
O.~Schwartz.
\newblock Untold history of {AI}: Algorithmic bias was born in the 1980s.
\newblock {\em IEEE Spectrum}, 2019.

\bibitem{shields2018over}
J.~Shields.
\newblock Over 98\% of fortune 500 companies use applicant tracking systems
  (ats), 2018.

\bibitem{smedinghoff2007art}
G.~Smedinghoff.
\newblock The art, philosophy and science of data.
\newblock {\em Contingencies May/June}, pages 37--40, 2007.

\bibitem{sobol1988measures}
M.~G. Sobol and C.~J. Ellard.
\newblock Measures of employment discrimination: A statistical alternative to
  the four-fifths rule.
\newblock {\em Industrial Relations Law Journal}, pages 381--399, 1988.

\bibitem{steele}
C.~M. Steele and J.~Aronson.
\newblock Stereotype threat and the intellectual test performance of african
  americans.
\newblock {\em Journal of Personality and Social Psychology}, 69:797--811,
  1995.

\bibitem{sullivan2005disparate}
C.~A. Sullivan.
\newblock Disparate impact: Looking past the desert palace mirage.
\newblock {\em William \& Mary Law Review}, 2005.

\bibitem{npr-article}
N.~Totenberg.
\newblock Supreme court delivers major victory to lgbtq employees, 2020.

\bibitem{wang2019your}
J.~Wang and N.~Shah.
\newblock Your 2 is my 1, your 3 is my 9: Handling arbitrary miscalibrations in
  ratings.
\newblock In {\em AAMAS Conference proceedings}, 2019.

\bibitem{wang2020debiasing}
J.~Wang, I.~Stelmakh, Y.~Wei, and N.~B. Shah.
\newblock Debiasing evaluations that are biased by evaluations.
\newblock {\em arXiv preprint arXiv:2012.00714}, 2020.

\bibitem{wold1997nepotism}
A.~Wold and C.~Wenner{\aa}s.
\newblock Nepotism and sexism in peer review.
\newblock {\em Nature}, 387(6631):341--343, 1997.

\bibitem{yucer2020exploring}
S.~Yucer, S.~Ak{\c{c}}ay, N.~Al-Moubayed, and T.~P. Breckon.
\newblock Exploring racial bias within face recognition via per-subject
  adversarially-enabled data augmentation.
\newblock In {\em Proceedings of the IEEE/CVF Conference on Computer Vision and
  Pattern Recognition Workshops}, pages 18--19, 2020.

\bibitem{zemel2013learning}
R.~Zemel, Y.~Wu, K.~Swersky, T.~Pitassi, and C.~Dwork.
\newblock Learning fair representations.
\newblock In {\em International Conference on Machine Learning}, pages
  325--333, 2013.

\end{thebibliography}

%%
%% If your work has an appendix, this is the place to put it.
\appendix

\section{Constructing the Poset}
\label{app:constructing-the-poset}

As discussed in Section \ref{sec:new-directions}, a \emph{poset} is a partially ranked set of candidates, and the \emph{poset approach} is the process of forming a partial ranking and subsequently making selections based on the ranking. The purpose of this approach is to minimize the effect of bias and other inaccuracies on selection decisions. In this section, we discuss several ways in which posets can be constructed.

One natural way to construct a partial ranking of applicants is to first form confidence intervals around raw scores, and then extract ordinal information from non-intersecting intervals (see, e.g., Examples \ref{ex:gender} and \ref{ex:2d}). There are several factors that can be taken into account when forming a partial ranking in such a way. An evaluation metric itself can produce inaccurate scores, which can inform the lengths of intervals. Group-specific biases and error rates can also inform varying interval lengths by group. Evaluation metrics can also be biased against some groups, and intervals can be designed so as to mitigate known biases. To illustrate some ways in which the poset approach can be implemented, we consider an example from \cite{closingthegap}.

\paragraph{Example case study.} In \cite{closingthegap}, Salem and Gupta analyzed the Aspiring Minds dataset, in which male and female job seekers had similar distributions of computer science test scores. As the job seekers were all in computer science fields, computer science test scores were taken as proxies for hireability. They found that by performing a linear regression on features (including gender and test scores), female\footnote{All entries in the dataset listed a gender of male or female.} applicants received scores that were 16.95 points lower than those of male applicants with all other attributes equal. %were penalized by 16.95 points due to their gender. 
This does not mean that the regression model will underestimate the score of every female; rather, it means that the regression model has ``learned'' a trend in the data that will be harmful to some female applicants.\footnote{One of the reasons for this might be imbalance of data points in the male versus female group. For example, the average score in the training set among all female applicants was 455.30, and the average score among all male applicants was 478.05 (a slightly larger discrepancy than in the entire dataset).}
Since the available data on applicants indicated that mean scores were similar across gender, this learned trend can be viewed as a bias which should be corrected. Importantly, even if gender were scrubbed from the data, a machine learning model might still pick up on these trends via proxy variables, so scrubbing information is in general not advisable \cite{kleinberg2019simplicity,dwork}.

Given that there is some data available on candidates, and a prediction mechanism that scores candidates, how can a partial ranking of the candidates be constructed in a mathematically sound way?\footnote{In the example above, the scoring mechanism was a simple linear regression. But in general, it can involve natural language processing of the text in r\'esum\'es, a neural network built on top of that, or any other machine learning technique.} In what follows, we outline three possible approaches for constructing ranges of scores for each candidate. However, we note that other statistical approaches might be applicable as well.\\

1. {\bf Data-centric approach to mitigate disparate error rates:} It is desirable that as the errors in evaluations decrease, candidates' score ranges shrink as well. This is statistically linked to the amount of data in the following way: a machine learning model trained on noisy data tends to have smaller errors as the amount of data increases. Therefore, in this approach, we will ensure that the size of the range decreases as the amount of the training data for a candidate increases. Formally, one can set the interval length to be $r(n)$, where $n$ is the number of points in the training data and $r$ some function that decays with $n$. The value of $r(n)$ should be an estimate of how much error is in the model after observing $n$ training points, which can be calculated depending on the model class and the extent of noise; for example, when the true relationship is linear, it makes sense to choose $r(n)$ to be proportional to $1/\sqrt{n}$ \cite{christensen1996analysis}.\footnote{If the true relationship between a candidate's features and hireability is linear, and the noise is independently drawn from identical normal distributions, then the linear regression estimation of hireability is unbiased and has standard deviation proportional to $1/\sqrt{n}$. The proportionality constant depends on the distribution of the datapoints and the variance of the noise \cite{christensen1996analysis}.} Adopting these confidence intervals gives applicants the benefit of the doubt, giving room to account for imperfections in the evaluation mechanism.
    
    As noted earlier, however, error rates can differ by group. The presence of an imbalanced training dataset (that is, one with an underrepresented group) can lead to higher error rates for the underrepresented group. This problem mathematically justifies having intervals of different lengths for different groups, dependent on how much uncertainty exists within each group. For example, if there were $n_1$ applicants from one group and $n_2$ applicants from another group in the training data, then one might choose intervals of length $r(n_1)$ for the former and intervals of length $r(n_2)$ for the latter. 

    In the context of the example case study, this would entail partitioning the pool into some number of groups and setting differing interval lengths dependent on group size. Since a bias was detected against female candidates, it may make sense to consider groups by gender, in which case the interval length for a female candidate would be proportional to $\frac{1}{\sqrt{\#\mbox{females}}}$, and analogously for men. There are, however, other ways to partition the pool. Each person is mapped to a point in some high-dimensional space, and clusters can be formed based on which points are close to each other with respect to some metric. We can then similarly set interval lengths based on cluster sizes.

If one insists on using equal interval lengths for all applicants, then using a length of, say, $\min\{r(n_1),r(n_2)\}$, would give each applicant the length corresponding to the error rate in the largest group. While this last method would arguably give overly narrow intervals to members of the smaller group, it still gives more benefit of the doubt to applicants (regardless of group) than using intervals of length $r(n)$. \\

2. {\bf Data-centric approach to mitigate group-specific biases:} When there is a known bias against groups of individuals (either due to the evaluation metric or some other systemic cause), intervals can be designed in a targeted way to correct for these errors. This might involve individualized interval lengths and translations to account for observed or inferred bias. Returning to the example case study, we will outline how Salem and Gupta accounted for the observed bias discussed above. They had hypothesized a multiplicative bias in the female group (as is assumed in the group model of bias of Kleinberg and Raghavan \cite{kr18}), meaning that a ``true'' ability $s^*$ of a female applicant would produce a score of $s^*/\beta$, for some bias factor $\beta$. They therefore sought to convert this additive error of 16.95 into a multiplicative error. Note that an error of 16.95 for a low-scoring individual produces a larger multiplicative error than does an additive error of 16.95 on a high-scoring individual. They therefore found upper and lower estimates on the multiplicative bias factor by observing the multiplicative bias factor of a low-scoring female applicant (mean $-$0.5 standard deviations) and a high-scoring female applicant (mean $+$0.5 standard deviations). The parameter 0.5 for standard deviation was chosen so that corresponding score ranges of female job seekers were (1) small enough that comparisons could still be made between enough individuals, and (2) large enough that bias was mitigated.\footnote{The process of tuning such a parameter is an important step in the design of a task-specific algorithm, and often involves trial-and-error experimentation on training data.} Once a lower multiplicative bias factor $\beta_{\text{low}}$ and an upper multiplicative bias fact $\beta_{\text{high}}$ were calculated, the interval for a female applicant of score $s$ would be $[s\beta_{\text{low}},s\beta_{\text{high}}]$. Using these intervals instead of raw scores gives female candidates more ``benefit of the doubt,'' which the authors considered appropriate given the observed bias. Ultimately, they showed that this method mitigated gender bias more effectively than a simpler gender-aware algorithm.

If feedback is available on previously evaluated applicants, then simpler approaches may be appropriate. For example, if it is known that for a particular group, the evaluation metric has a standard deviation of 1 and tends to underestimate scores by two points, then a score of 83 in that group could justifiably be translated to an interval of $[84,86]$.\\

3. {\bf Human-centric approach:}
     When there is direct human involvement in evaluating applicants (e.g., through interviews), biases and inaccuracies might still persist. One might think that the poset approach would not help in this situation, but it can. Similar to the data-centric approach, one can reduce this error through repeated independent evaluations (i.e., wisdom of crowds \cite{smedinghoff2007art}). In this case, a single applicant can be interviewed by a diverse committee, each of whose members scores the applicant. These scores can then be used to form score ranges for each applicant (e.g., the minimum score to the maximum score, or the first to the third quartile, etc.). An applicant's score range can be decreased through further discussion by the committee until the interval is sufficiently small to allow for reasonably many comparisons in the applicant pool.\footnote{Such practices are already prevalent within hiring committees and program committees for conferences such as ICLR, NeurIPS and WWW.} If there were some way to obtain repeated evaluations (e.g., through multiple human evaluators, or access to multiple scoring algorithms) of the job seekers in the example case study, then this human-centric approach could be applied there as well. For each person in the dataset, we could repeatedly obtain evaluations until we can be confident in their score (say, up to 5 points). If certain groups (e.g., gender groups, or clusters obtained by a machine learning algorithm) experience higher variance in evaluations, then the same number of evaluations might result in confidence intervals of different lengths for different groups.

\paragraph{Accounting for other sources of errors.} There are sources of errors which are intrinsic to machine learning and predictions, and interval lengths can also be designed to mitigate these errors. For example, if a highly informative or causal variable is absent from the data, then predictions can suffer (this type of error is called \emph{Bayes error}). Another type of error, called \emph{approximation error}, describes error resulting from the mismatch between the true relationship between applicant features and ability, and the class of models that can be produced by the algorithm. Both of these types of errors are independent of the size of the training data, so we generally do not expect these errors to decay with time. In an attempt to account for this sort of intrinsic error, one could impose a minimum interval length for all candidates, where the minimum length depends on the accuracy of the evaluation metric.

\begin{figure}[t]
    \begin{minipage}[c]{0.43\textwidth}
    \centering
    \includegraphics[width=\textwidth]{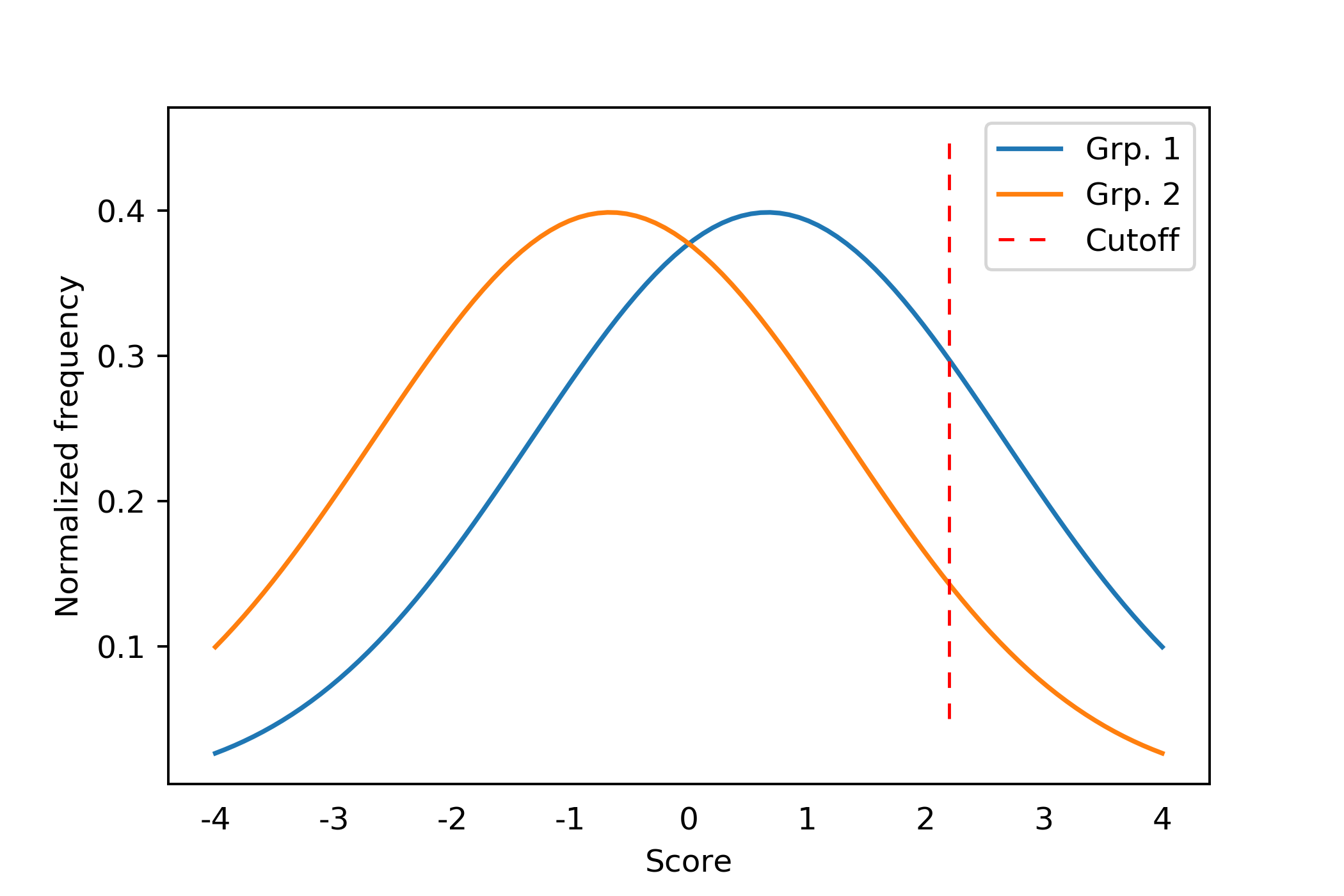}
    \end{minipage}
    \begin{minipage}[c]{0.55\textwidth}
    \centering 
    \scalebox{.7}{
    \begin{tikzpicture}
    
    \def\b{.75}
    \def\o{1}
    
    \tikzstyle{vertex}=[circle, draw, fill = blue, fill opacity = .1, text opacity = 1]
        \draw[dashed,<->] (-5,-.7) -- (5,-.7);
    \foreach \a in {-4,...,4}{
    \draw (\a,-.8) -- (\a,-.6);
    \node at (\a,-1.08){\a};
    }

    \draw[ultra thick,newblue] (3-\b,-.3) -- (3+\b,-.3);
    \node at (3+\b+.3,-.2){\textcolor{newgreen}{$\checkmark$}};
    \draw[ultra thick,newblue] (2.5-\b,-.1) -- (2.5+\b,-.1);
    \node at (2.5+\b+.3,0){\textcolor{newgreen}{$\checkmark$}};
    \draw[ultra thick,newblue] (-2.3-\b,-.1) -- (-2.3+\b,-.1);
    \draw[ultra thick,newblue] (-2.8-\b,.5) -- (-2.8+\b,.5);
    \draw[ultra thick,neworange] (2.1-\o,.5) -- (2.1+\o,.5);
    \node at (2.1+\o+.3,.6){\textcolor{newgreen}{$\checkmark$}};
    \draw[ultra thick,newblue] (1.8-\b,.1) -- (1.8+\b,.1);
    \node at (1.8+\b+.3,.2){\textcolor{newgreen}{$\checkmark$}};
    \draw[ultra thick,newblue] (1-\b,.3) -- (1+\b,.3);
    \draw[ultra thick,newblue] (.7-\b,-.3) -- (.7+\b,-.3);
    \draw[ultra thick,newblue] (-\b,-.1) -- (\b,-.1);
    \draw[ultra thick,neworange] (-.6-\o,.5) -- (-.6+\o,.5);
    \draw[ultra thick,neworange] (-.8-\o,.1) -- (-.8+\o,.1);
    \draw[ultra thick,newblue] (-1.6-\b,.3) -- (-1.6+\b,.3);
    \draw[ultra thick,neworange] (-2-\o,-.3) -- (-2+\o,-.3);
    \draw[ultra thick,newblue] (-3.1-\b,.1) -- (-3.1+\b,.1);
    \draw[ultra thick,newblue] (-\b,.7) -- (\b,.7);
    
    \draw[dashed,<->] (-5,2.6) -- (5,2.6);
    \foreach \a in {-4,...,4}{
    \draw (\a,2.5) -- (\a,2.7);
    \node at (\a,2.22){\a};
    }
    \tikzstyle{vertex}=[regular polygon,regular polygon sides=3, draw, color = newblue, fill = newblue, text opacity = 1, inner sep = 1.8pt]
    \node[vertex](b2) at (1.8,3.5){};
    \node[vertex](b3) at (1,3.7){};
    \node[vertex](b4) at (.7,3.1){};
    \node[vertex](b5) at (0,3.3){};
    \node[vertex](b7) at (-.8,3.5){};
    \node[vertex](b8) at (-1.6,3.7){};
    \node[vertex](b10) at (-3.1,3.5){};
    \node[vertex](b11) at (0,4.1){};
    \node[vertex](o1) at (3,3.1){}; %add 3.4
    \node at (3.3,3.2){\textcolor{newgreen}{$\checkmark$}};
    \node[vertex](o2) at (2.5,3.3){};
    \node at (2.8,3.4){\textcolor{newgreen}{$\checkmark$}};
    \node[vertex](o3) at (-2.3,3.3){};
    \node[vertex](o4) at (-2.8,3.9){};
    
    \tikzstyle{vertex}=[regular polygon,regular polygon sides=3, draw, color = neworange, fill = neworange, text opacity = 1, inner sep = 1.8pt]
    
    \node[vertex](b9) at (-2,3.1){};
    \node[vertex](b7) at (-.8,3.5){};
    \node[vertex](b6) at (-.6,3.9){};
    \node[vertex](b1) at (2.1,3.9){};
    
    \draw[thick,dashed,red] (2.3,-2) -- (2.3,4.9);
    \node at (3,4.6){\textcolor{red}{Cutoff}};
    
    \node at (0,-1.5){Adjusted score ranges};
    \node at (0,1.8){Raw scores};
    \end{tikzpicture}
    }
    \end{minipage}
    \caption{\footnotesize (left) Example of score distributions (blue: Group 1, orange: Group 2) and (right) potential score ranges for candidates from these distributions. Suppose a hiring committee wants to select two of the applicants represented in the right plot. If only the raw evaluations (the centers of the intervals) are used to make these decisions, then only the two high-scoring Group 1 candidates could be selected, as they are the only applicants meeting the cutoff. However, if score ranges are considered, then the highest-scoring Group 2 candidate meets the cutoff as well. In this example, adopting the poset method results in a more diverse slate of candidates meeting the cutoff, vis-\`a-vis using raw scores.
    }
    \label{fig:poset-not-quotas-unimodal}
\end{figure}
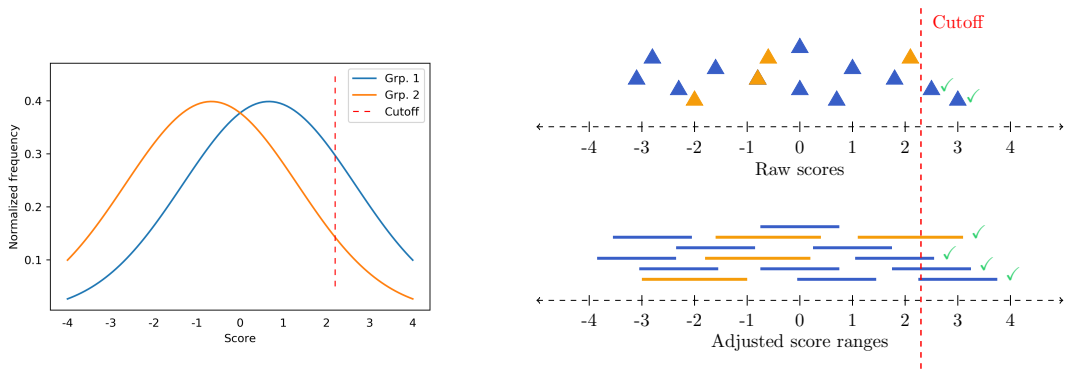

\begin{figure}[t]
    \begin{minipage}[c]{0.43\textwidth}
    \centering
    \includegraphics[width=\textwidth]{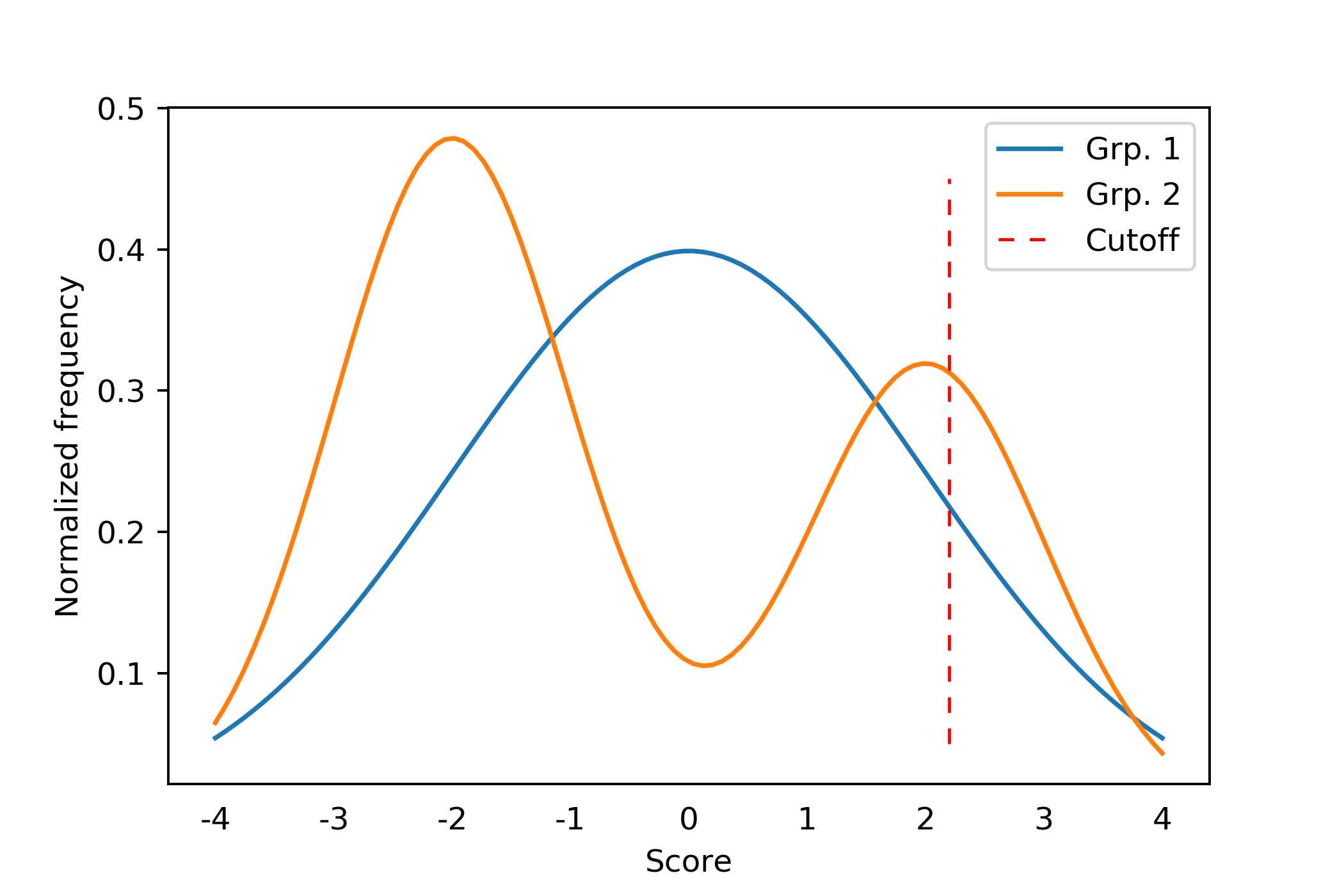}
    \end{minipage}
    \begin{minipage}[c]{0.55\textwidth}
    \centering
    \scalebox{.7}{
    \begin{tikzpicture}
    
    \def\b{.75}
    \def\o{1}
    
    \tikzstyle{vertex}=[circle, draw, fill = blue, fill opacity = .1, text opacity = 1]
        \draw[dashed,<->] (-5,-.7) -- (5,-.7);
    \foreach \a in {-4,...,4}{
    \draw (\a,-.8) -- (\a,-.6);
    \node at (\a,-1.08){\a};
    }

    \draw[ultra thick,neworange] (3 - \o,-.3) -- (3+\o,-.3);
    \node at (3+\o+.3,-.2){\textcolor{newgreen}{$\checkmark$}};
    \draw[ultra thick,neworange] (2.5-\o,-.1) -- (2.5+\o,-.1);
    \node at (2.5+\o+.3,0){\textcolor{newgreen}{$\checkmark$}};
    \draw[ultra thick,neworange] (-2.3-\o,-.1) -- (-2.3+\o,-.1);
    \draw[ultra thick,neworange] (-2.8-\o,.5) -- (-2.8+\o,.5);
    \draw[ultra thick,newblue] (2.1-\b,.5) -- (2.1+\b,.5);
    \node at (2.1+\b+.3,.6){\textcolor{newgreen}{$\checkmark$}};
    \draw[ultra thick,newblue] (1.8-\b,.1) -- (1.8+\b,.1);
    \node at (1.8+\b+.3,.2){\textcolor{newgreen}{$\checkmark$}};
    \draw[ultra thick,newblue] (1-\b,.3) -- (1+\b,.3);
    \draw[ultra thick,newblue] (.7-\b,-.3) -- (.7+\b,-.3);
    \draw[ultra thick,newblue] (-\b,-.1) -- (\b,-.1);
    \draw[ultra thick,newblue] (-.6-\b,.5) -- (-.6+\b,.5);
    \draw[ultra thick,newblue] (-.8-\b,.1) -- (-.8+\b,.1);
    \draw[ultra thick,newblue] (-1.6-\b,.3) -- (-1.6+\b,.3);
    \draw[ultra thick,newblue] (-2-\b,-.3) -- (-2+\b,-.3);
    \draw[ultra thick,newblue] (-3.1-\b,.1) -- (-3.1+\b,.1);
    \draw[ultra thick,newblue] (-\b,.7) -- (\b,.7);
    
    \draw[dashed,<->] (-5,2.6) -- (5,2.6);
    \foreach \a in {-4,...,4}{
    \draw (\a,2.5) -- (\a,2.7);
    \node at (\a,2.22){\a};
    }
    \tikzstyle{vertex}=[regular polygon,regular polygon sides=3, draw, color = newblue, fill = newblue, text opacity = 1, inner sep = 1.8pt]
    \node[vertex](b1) at (2.1,3.9){};
    \node[vertex](b2) at (1.8,3.5){};
    \node[vertex](b3) at (1,3.7){};
    \node[vertex](b4) at (.7,3.1){};
    \node[vertex](b5) at (0,3.3){};
    \node[vertex](b6) at (-.6,3.9){};
    \node[vertex](b7) at (-.8,3.5){};
    \node[vertex](b8) at (-1.6,3.7){};
    \node[vertex](b9) at (-2,3.1){};
    \node[vertex](b10) at (-3.1,3.5){};
    \node[vertex](b11) at (0,4.1){};

    \tikzstyle{vertex}=[regular polygon,regular polygon sides=3, draw, color = neworange, fill = neworange, text opacity = 1, inner sep = 1.8pt]
    \node[vertex](o1) at (3,3.1){}; %add 3.4
    \node at (3.3,3.2){\textcolor{newgreen}{$\checkmark$}};
    \node[vertex](o2) at (2.5,3.3){};
    \node at (2.8,3.4){\textcolor{newgreen}{$\checkmark$}};
    \node[vertex](o3) at (-2.3,3.3){};
    \node[vertex](o4) at (-2.8,3.9){};

    \draw[thick,dashed,red] (2.3,-2) -- (2.3,4.9);
    \node at (3,4.6){\textcolor{red}{Cutoff}};
    
    \node at (0,-1.5){Adjusted score ranges};
    \node at (0,1.8){Raw scores};
    \end{tikzpicture}
    }
    \end{minipage}
    \caption{(left) Example of (somewhat unusual) score distributions (blue: Group 1, orange: Group 2) and (right) potential score ranges for candidates from these distributions. Suppose a hiring committee wants to select two of the applicants corresponding to the right plot. If only the raw evaluations (the centers of the intervals) are used to make these decisions, then only the two high-scoring Group 2 candidates could be selected, as they are the only applicants meeting the cutoff. However, if score ranges are considered, then the two highest-scoring Group 1 candidates meet the cutoff as well. From this example, we see that adopting the poset approach can be beneficial to the majority group as well and does not routinely advantage the lower-mean group (in this case, Group 2). 
    }
    \label{fig:poset-not-quotas-bimodal}
\end{figure}

\begin{figure}[t]
    \begin{minipage}[c]{0.43\textwidth}
    \centering
    \includegraphics[width=\textwidth]{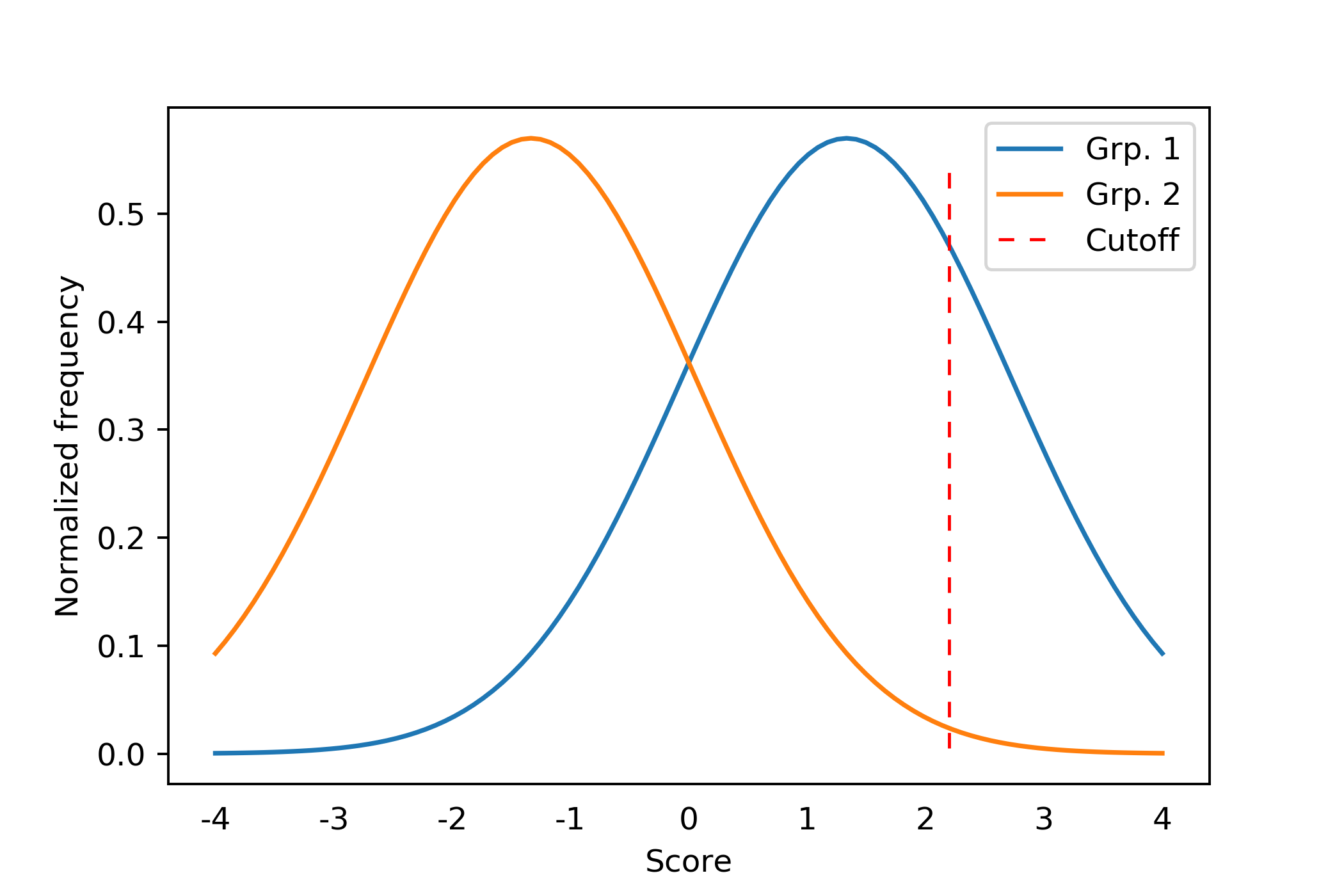}
    \end{minipage}
    \begin{minipage}[c]{0.55\textwidth}
    \centering
    \scalebox{.7}{
    \begin{tikzpicture}
    
    \def\b{.75}
    \def\o{1}
    
    \tikzstyle{vertex}=[circle, draw, fill = blue, fill opacity = .1, text opacity = 1]
        \draw[dashed,<->] (-5,-.7) -- (5,-.7);
    \foreach \a in {-4,...,4}{
    \draw (\a,-.8) -- (\a,-.6);
    \node at (\a,-1.08){\a};
    }

    \draw[ultra thick,newblue] (3-\b,-.3) -- (3+\b,-.3);
    \node at (3+\b+.3,-.2){\textcolor{newgreen}{$\checkmark$}};
    \draw[ultra thick,newblue] (2.5-\b,-.1) -- (2.5+\b,-.1);
    \node at (2.5+\b+.3,0){\textcolor{newgreen}{$\checkmark$}};
    \draw[ultra thick,neworange] (-2.3-\o,-.1) -- (-2.3+\o,-.1);
    \draw[ultra thick,neworange] (-2.8-\o,.5) -- (-2.8+\o,.5);
    \draw[ultra thick,newblue] (2.1-\b,.5) -- (2.1+\b,.5);
    \node at (2.1+\b+.3,.6){\textcolor{newgreen}{$\checkmark$}};
    \draw[ultra thick,newblue] (1.8-\b,.1) -- (1.8+\b,.1);
    \node at (1.8+\b+.3,.2){\textcolor{newgreen}{$\checkmark$}};
    \draw[ultra thick,neworange] (1-\o,.3) -- (1+\o,.3);
    \draw[ultra thick,newblue] (.7-\b,-.3) -- (.7+\b,-.3);
    \draw[ultra thick,newblue] (-\b,-.1) -- (\b,-.1);
    \draw[ultra thick,newblue] (-.6-\b,.5) -- (-.6+\b,.5);
    \draw[ultra thick,newblue] (-.8-\b,.1) -- (-.8+\b,.1);
    \draw[ultra thick,neworange] (-1.6-\o,.3) -- (-1.6+\o,.3);
    \draw[ultra thick,newblue] (-2-\b,-.3) -- (-2+\b,-.3);
    \draw[ultra thick,newblue] (-3.1-\b,.1) -- (-3.1+\b,.1);
    \draw[ultra thick,newblue] (-\b,.7) -- (\b,.7);
    
    \draw[dashed,<->] (-5,2.6) -- (5,2.6);
    \foreach \a in {-4,...,4}{
    \draw (\a,2.5) -- (\a,2.7);
    \node at (\a,2.22){\a};
    }
    \tikzstyle{vertex}=[regular polygon,regular polygon sides=3, draw, color = newblue, fill = newblue, text opacity = 1, inner sep = 1.8pt]
    \node[vertex](b1) at (2.1,3.9){};
    \node[vertex](b2) at (1.8,3.5){};
    \node[vertex](b4) at (.7,3.1){};
    \node[vertex](b5) at (0,3.3){};
    \node[vertex](b6) at (-.6,3.9){};
    \node[vertex](b7) at (-.8,3.5){};
    \node[vertex](b9) at (-2,3.1){};
    \node[vertex](b10) at (-3.1,3.5){};
    \node[vertex](b11) at (0,4.1){};
    \node[vertex](o2) at (2.5,3.3){};
    \node[vertex](o1) at (3,3.1){};

    \tikzstyle{vertex}=[regular polygon,regular polygon sides=3, draw, color = neworange, fill = neworange, text opacity = 1, inner sep = 1.8pt] 
    \node at (3.3,3.2){\textcolor{newgreen}{$\checkmark$}};
    \node at (2.8,3.4){\textcolor{newgreen}{$\checkmark$}};
    \node[vertex](o3) at (-2.3,3.3){};
    \node[vertex](o4) at (-2.8,3.9){};
    \node[vertex](b3) at (1,3.7){};
    \node[vertex](b8) at (-1.6,3.7){};
    
    \draw[thick,dashed,red] (2.3,-2) -- (2.3,4.9);
    \node at (3,4.6){\textcolor{red}{Cutoff}};
    
    \node at (0,-1.5){Adjusted score ranges};
    \node at (0,1.8){Raw scores};
    \end{tikzpicture}
    }
    \end{minipage}
    \caption{(left) Example of score distributions (blue: Group 1, orange: Group 2) and (right) potential score ranges for candidates from these distributions. Suppose a hiring committee wants to select two of the applicants corresponding to the right plot. If only the raw evaluations (the centers of the intervals) are used to make these decisions, then only the two highest-scoring Group 1 candidates could be selected, as they are the only applicants meeting the cutoff. If the score ranges are considered, then the four highest-scoring Group 1 candidates meet the cutoff. This example shows that adopting the poset approach does not necessarily increase the selection rate for the group with the lower mean score, and that the poset approach does not necessarily constitute a quota system.
    }
    \label{fig:poset-not-quotas-separated}
\end{figure}

\begin{figure}[htbp]
    \begin{minipage}[c]{0.43\textwidth}
    \centering
    \includegraphics[width=\textwidth]{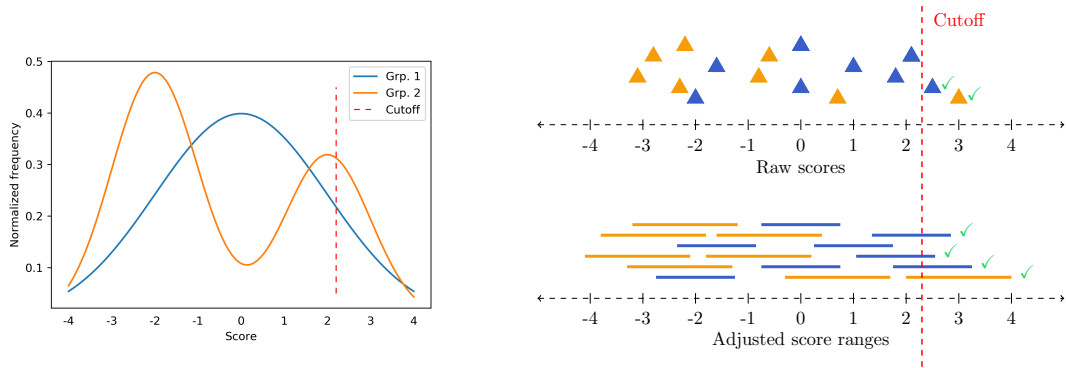}
    \end{minipage}
    \begin{minipage}[c]{0.55\textwidth}
    \centering
    \scalebox{.7}{
    \begin{tikzpicture}
    
    \def\b{.75}
    \def\o{1}
    
    \tikzstyle{vertex}=[circle, draw, fill = blue, fill opacity = .1, text opacity = 1]
        \draw[dashed,<->] (-5,-.7) -- (5,-.7);
    \foreach \a in {-4,...,4}{
    \draw (\a,-.8) -- (\a,-.6);
    \node at (\a,-1.08){\a};
    }
    \draw[ultra thick,neworange] (3-\o,-.3) -- (3+\o,-.3);
    \node at (3+\o+.3,-.2){\textcolor{newgreen}{$\checkmark$}};
    \draw[ultra thick,newblue] (2.5-\b,-.1) -- (2.5+\b,-.1);
    \node at (2.5+\b+.3,0){\textcolor{newgreen}{$\checkmark$}};
    \draw[ultra thick,neworange] (-2.3-\o,-.1) -- (-2.3+\o,-.1);
    \draw[ultra thick,neworange] (-2.8-\o,.5) -- (-2.8+\o,.5);
    \draw[ultra thick,newblue] (2.1-\b,.5) -- (2.1+\b,.5);
    \node at (2.1+\b+.3,.6){\textcolor{newgreen}{$\checkmark$}};
    \draw[ultra thick,newblue] (1.8-\b,.1) -- (1.8+\b,.1);
    \node at (1.8+\b+.3,.2){\textcolor{newgreen}{$\checkmark$}};
    \draw[ultra thick,newblue] (1-\b,.3) -- (1+\b,.3);
    \draw[ultra thick,neworange] (.7-\o,-.3) -- (.7+\o,-.3);
    \draw[ultra thick,newblue] (-\b,-.1) -- (\b,-.1);
    \draw[ultra thick,neworange] (-.6-\o,.5) -- (-.6+\o,.5);
    \draw[ultra thick,neworange] (-.8-\o,.1) -- (-.8+\o,.1);
    \draw[ultra thick,newblue] (-1.6-\b,.3) -- (-1.6+\b,.3);
    \draw[ultra thick,newblue] (-2-\b,-.3) -- (-2+\b,-.3);
    \draw[ultra thick,neworange] (-3.1-\o,.1) -- (-3.1+\o,.1);
    \draw[ultra thick,newblue] (-\b,.7) -- (\b,.7);
    \draw[ultra thick,neworange] (-2.2-\o,.7) -- (-2.2+\o,.7);
    
    \draw[dashed,<->] (-5,2.6) -- (5,2.6);
    \foreach \a in {-4,...,4}{
    \draw (\a,2.5) -- (\a,2.7);
    \node at (\a,2.22){\a};
    }
    \tikzstyle{vertex}=[regular polygon,regular polygon sides=3, draw, color = newblue, fill = newblue, text opacity = 1, inner sep = 1.8pt]
    \node[vertex](b1) at (2.1,3.9){};
    \node[vertex](b3) at (1,3.7){};
    \node[vertex](b5) at (0,3.3){};
    \node[vertex](b8) at (-1.6,3.7){};
    \node[vertex](b9) at (-2,3.1){};
    \node[vertex](b11) at (0,4.1){};
    \node[vertex](o2) at (2.5,3.3){};
    \node[vertex](b1) at (2.1,3.9){};
    \node[vertex](b2) at (1.8,3.5){};
    \node[vertex](b11) at (0,4.1){};
    
    \tikzstyle{vertex}=[regular polygon,regular polygon sides=3, draw, color = neworange, fill = neworange, text opacity = 1, inner sep = 1.8pt] %add 3.4
    \node at (3.3,3.2){\textcolor{newgreen}{$\checkmark$}};
    \node at (2.8,3.4){\textcolor{newgreen}{$\checkmark$}};
    \node[vertex](o3) at (-2.3,3.3){};
    \node[vertex](o4) at (-2.8,3.9){};
    \node[vertex](o1) at (3,3.1){};
    \node[vertex](b7) at (-.8,3.5){};
    \node[vertex](b6) at (-.6,3.9){};
    \node[vertex](b4) at (.7,3.1){};
    \node[vertex](b11) at (-2.2,4.1){};
    \node[vertex](b10) at (-3.1,3.5){};

    \draw[thick,dashed,red] (2.3,-2) -- (2.3,4.9);
    \node at (3,4.6){\textcolor{red}{Cutoff}};
    
    \node at (0,-1.5){Adjusted score ranges};
    \node at (0,1.8){Raw scores};
    \end{tikzpicture}
    }
    \end{minipage}
    \caption{(left) Example of (somewhat unusual) score distributions (blue: Group 1, orange: Group 2) and (right) potential score ranges for candidates from these distributions. Suppose a hiring committee wants to select two of the applicants corresponding to the right plot. If only the raw evaluations (the centers of the intervals) are used to make these decisions, then one Group 1 and one Group 2 candidate will be selected, as they are the only applicants meeting the cutoff. In this case, demographic parity is achieved, as both groups have equal size. However, if score ranges are considered, then two additional Group 1 candidates meet the cutoff as well. This example shows that adopting the poset approach does not necessarily make the new candidate slate (i.e., those meeting the cutoff) more representative compared to using raw scores---indeed, in this example, adopting the poset approach moves the new candidate slate farther away from demographic parity.
    }
    \label{fig:poset-not-quotas-bimodal-nonmonotonic}
\end{figure}

\section{Poset Approach Diagrams}
\label{app:poset-not-quotas}

In Section \ref{sec:new-directions}, we discussed how the poset approach can be used in the screening of applications. To determine the appropriateness of this approach to screening, it is of legal, ethical, and utilitarian importance to understand the effect of the poset approach on applicants. While the practical effect of the poset approach will depend on context and precise implementation, it is informative to observe its effect on artificial data.

The four examples shown in this section (Figures \ref{fig:poset-not-quotas-unimodal}-\ref{fig:poset-not-quotas-bimodal-nonmonotonic}) compare candidate slates produced by a cutoff on raw scores versus on score ranges. To illustrate the effect of using score ranges over groups, we consider two groups, where Group 2 has a lower mean evaluation and a larger interval length (perhaps due to dataset imbalance issues as discussed earlier). In Figure \ref{fig:poset-not-quotas-unimodal}, we see how using score ranges instead of raw scores can increase the selection rate of the minority group. However, the use of score ranges does not necessarily benefit the minority group in general. Figures \ref{fig:poset-not-quotas-bimodal}-\ref{fig:poset-not-quotas-bimodal-nonmonotonic} consider the same scenario as Figure \ref{fig:poset-not-quotas-unimodal} but with different distributional assumptions. Of note is how the use of score ranges can benefit the majority or minority group, and can benefit the low-scoring or the high-scoring group. This point, in particular, means that the poset approach does not inherently constitute a quota system, which is an important feature with respect to anti-discrimination law (see Section \ref{subsec:corrective-action}).

\end{document}